\documentclass[final,reqno,conference]{IEEEtran} 
\usepackage{amsmath,amssymb,amsthm} \usepackage{graphicx,color} 

\title{Key technologies to accelerate the ICT Green evolution \\ 
\begin{Large} 
  An operator's point of view
\end{Large} 
\title{An operator's point of view}}
\author{\IEEEauthorblockN{Azeddine Gati, Fatma Ezzahra Salem, Ana Maria Galindo Serrano,  \\ Didier Marquet, Stephane Le Masson, Thomas Rivera, \\ Dinh-Thuy Phan-Huy, Zwi Altman, Jean-Baptiste Landre, \\ Olivier Simon, Esther Le Rouzic, Fabrice Bourgart, \\ Stephane Gosselin, Marc Vautier, Eric Gourdin, Taoufik En-Najjary, \\ Mamdouh El-Tabach, Raluca-Maria Indre, Guillaume Gerard, Gwenaelle Delsart}
\IEEEauthorblockA{Orange Labs, Ch\^atillon, France 
\\Email: \{fatma.salem,anamaria.galindoserrano,azeddine.gati\}@orange.com}\vspace{0.01cm}}

\begin{document}

\maketitle


\begin{abstract} 
The exponential growth in networks' traffic accompanied by the multiplication of new services like those promised by the 5G led to a huge increase in the infrastructures' energy consumption. All over the world, many telecom operators are facing the problem of energy consumption and Green networking since many years and they all convey today that it turned from sustainable development initiative to an OPEX issue. Therefore, the challenge to make the ICT sector more energy-efficient and environment-friendly has become a fundamental objective not only to green networks but also in the domain of green services that enable the ICT sectors to help other industrial sector to clean their own energy consumption.  
The present paper is a point of view of a European telecom operator regarding green networking. We address some technological advancements that would enable to accelerate this ICT green evolution after more than 15 years of field experience and international collaborative research projects. Basically, the paper is a global survey of the evolution of the ICT industry in green networks including optical and wireless networks and from hardware improvement to the software era as well as the green orchestration. 
\\


\end{abstract}

\section{Introduction}
There is no doubt on the role played by the Information and Communication Technology (ICT) sector  in economy, making cities more efficient and bringing communication everywhere. Nevertheless, the sector itself is responsible for between 2\% and 2.5\% of the total global carbon emissions \cite{ref1} and has important tendencies to continue growing.

\begin{figure} [ht]
\centering
\includegraphics[scale=0.5]{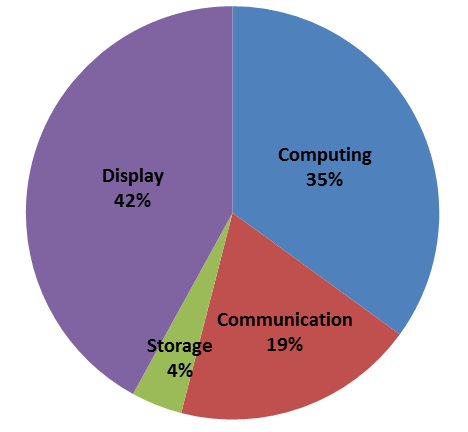}
\caption {Energy consumption functionalities shares of ICT systems.} 
\label {EC_share}
 \end{figure}

The four main information processing functions in modern ICT-electronic systems are: computation, communication, storage, and display, as shown in Figure \ref{EC_share} \cite{ref2}. Particularly, today energy costs for large service providers are substantial, and in the case of communication networks, can be a large component of Operational Expenditures (OPEX).

Recent studies of network energy use have shown that the power consumption today is dominated by the access equipment, i.e., edge points, although the transport network equipment will remain stable even with increasing traffic.

Reference \cite{ref3} shows the evolution until 2020 of the total power-per-user across all services for the Business as Usual (BAU) trends. We can observe that the energy use pattern for the different network components increases due to the expected traffic growth. This power increase is due to the fact that the rate of efficiency improvement is slower than the traffic growth rate. Particularly, for the fixed access the power remains almost constant since authors assume the use of Optical Line Termination (OLT), whose energy consumption remains somehow constant independently of the load.

It is true that the energy consumption issue has always been part of the environmental footprint and social responsibility. Nowadays, the concern on energy consumption diversifies and other disciplines such as systems engineering, applied research or marketing, are strongly working on this subject, especially with the emergence of the energy market and the smart grids.

The green subject is transverse and encouraged by the industry, the public authorities and by regulators. For instance, multiple projects were launched during the last years such as GreenTouch \cite{GreenT}, FP7-Earth \cite{Earth}, SoGreen \cite{SoGreen}, Opera-Net \cite{OpNet}, among which many have been funded by the European community. 

Sustainable development is a very broad area. For a long time it has been confined to the “life sciences”, a new branch which is taught from the primary education. Due to the magnitude of the climate, the ozone layer and waste management issues, the sustainable development topic has drifted, under the pressure of certain lobbies, to purely regulatory or societal aspects such as negative growth \cite{ref4}. The subject is still perceived as an additional tax or expense. In the ICT sector, the sustainable development topic was quickly transformed into an opportunity for growth and cost reduction as soon as it was identified by other sectors as a major axis enabling to reduce their environmental impact.

The digital information transport requires networks that basically consist of transmitters and receivers exchanging information via media such as copper, fiber optics or the air (wireless). At the end of the chain, we find either client devices in the one hand, servers or Data Centers (DCs) on the other hand. To convey information from one point to another, each network element consumes locally electrical energy. Ideally, the equipment energy consumed should be equal to the energy transmitted. However, this is not the case, mainly because of the heat losses in electronic circuits, the execution of local treatments (modulation, conversion, Field-Programmable Gate Array (FPGA), etc) or even due to the communication protocols. 

\begin{figure} [ht]
\centering
\includegraphics[width=6cm]{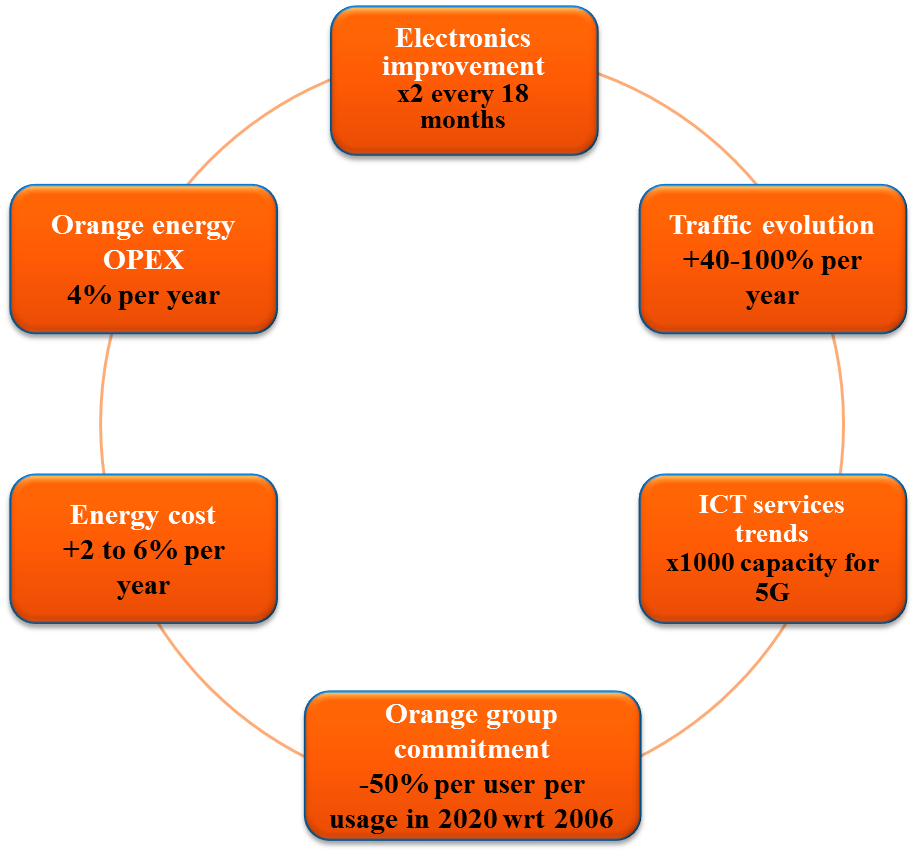}
\caption {Orange group green challenges} 
\label {fig:greenchallenge}
 \end{figure}

The ``green" challenge is directly related to the exponential growth of the network traffic. Orange group expects an annual increase in traffic of about 40\% to 100\% per year, while we observe a decrease in the Average Revenue Per User (ARPU) due to the highly competitive market on the one hand, especially in developed countries, and on the other hand, due to operators increasing OPEX investment such as the energy growing cost. Figure \ref{fig:greenchallenge} gives a summary of the constraints for a telecom group like Orange.

Even though the evolution of electronics allows a favorable trend in energy consumption, it could be completely absorbed by the introduction of new services such as High-Definition Television (HDTV), gaming, and 3D applications which require huge bandwidth and energy resources.

From a global network point of view, the energy consumption is about 10 W/user in access, 1 W/user in transport and 0,1 W/user in the core. These figures do not reflect the individual consumption of each network element, e.g., a server consumes much more than an access point or a Base Station (BS), but rather the number of elements in each sector and the number of clients connected to each network element. 

The challenge of reducing environmental footprint is shared by all the major operators in the world. Their stated objectives vary according to their networks and ecosystems characteristics.
The environmental footprint reduction targets vary from operator to operator and the used references are not the same. Certain operators have not even defined any reference date for their green transformation plan.
 
Note that the measurement index is also variable, e.g., some operators talk about greenhouse gas emission while others use energy consumption. However, all operators are now considering energy reduction as a major objective in their strategy because it is a constant growing expense. Future 5G design for example, has arised low energy consumption in the top 5 requirements list. 

We present in this paper some technological levers that would enable to enhance the Energy Efficiency (EE) in future ICT systems. For instance, simplification of the network's architecture, green orchestration, improvement of materials and electronics, optical scalability, etc. Many networks' segments are impacted by these advancements: access network, core network, switching/routing and data centers.


The remainder of this paper is as follows: Section \ref{section1} gives an insight into the fundamental laws for communication systems especially Shannon limit and Moore's law. Section \ref{section2} introduces the future trends in semiconductors, both in materials and components. Section \ref{section3} introduces the recent green advancements in optical networks. Section \ref{section4} presents various solutions enabling to enhance the energy efficieny in wireless networks. Section \ref{section5} presents the challenges behind the real time orchestration of the network under energy consumption constraints. Section \ref{section6} shows how to adapt the network's architecture towards a greener one. Section \ref{section7} presents the recent green studies on software eco-design, cooling and energy storage. Section \ref{section8} presents some mathematical approaches used for green for instance game theory for assessing services energy consumption and Reinforcement Learning for ASM orchestration in 5G networks. Finally Section \ref{section9} concludes the paper and gives some perspectives.   

\section{Fundamental laws for communications}
\label{section1}
This section is an overview of the fundamentals and laws that drive telecommunication system capacity and energy consumption prediction. 

\subsection{Shannon's law}
This law defines the limit of the capacity of a propagation channel between a simple telecommunication transmitter and a receiver. It sets out the maximum flow that can be achieved in a link and it is expressed as the link capacity:
\begin{equation}
	C \leq W*log⁡(1+\frac{P}{N})
\end{equation}
where $W$ is the signal bandwidth, $P$ is the received power at the receiver and $N$ is the noise including interference. The received power represents the amount of power induced by the transmitter and collected by the receptor after propagation.

This law (or rather limit) underlines a fundamental principle: to increase the capacity of a network, one must increase either the frequency band or the received power at the receiver. Increasing the frequency band is not a straightforward solution since bandwidth is very expensive, and also complicated from a hardware point view. That is why legacy networks high data rates have been essentially driven by an increase of the transmitted power.

From an EE point of view, the telecommunications equipment enhancement is continuous, around +2\% per year. Nevertheless, we observe that the increase in capacity comes together with a significant increase in the bandwidth used for these technologies. Indeed, in the case of the radio, we passed from 2G with a signal bandwidth of 200 kHz per carrier to 4G with up to 20 MHz band.

\begin{figure} [ht]
\centering
\includegraphics[width=6cm]{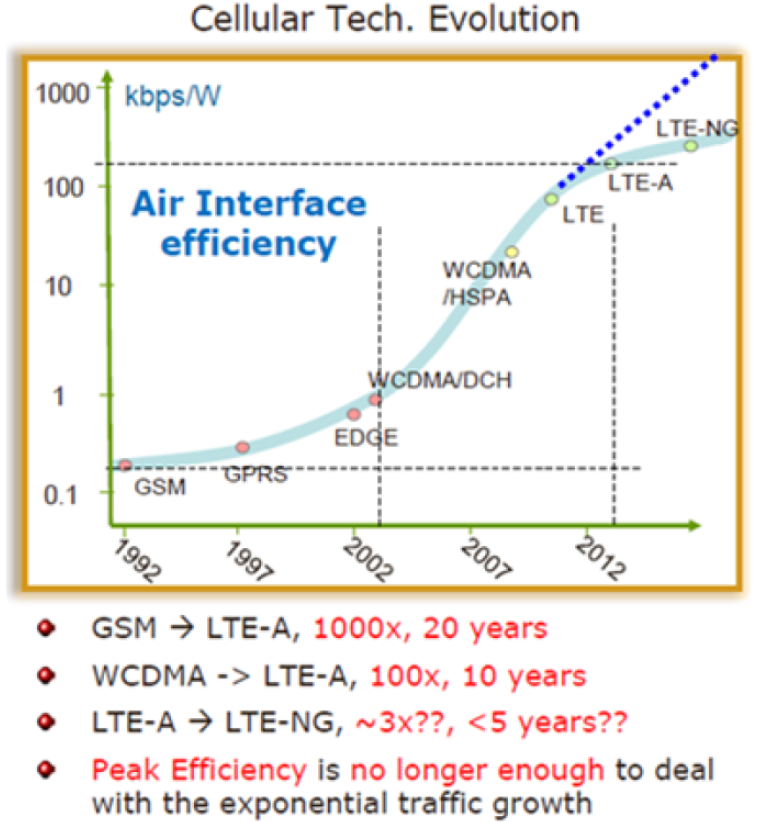}
\caption {BS transmission capacities per unit of energy} 
\label {BStx}
\end{figure}

We can see in Figure \ref{BStx} that over the last 20 years the BSs have multiplied their transmission capacity per unit of energy by 1000. The hidden problem behind this favorable development is that the rates boost has been essentially allowed by an increase in the transmitted power which induces more energy consumption. In any case, in the past the efforts have been mainly put in improving the systems Spectral Efficiency (SE) while now it is necessary to focus on the systems EE. Notice that the target of multiplying the EE by 1000 has been demonstrated during the GreenTouch project. 

Telecommunication operators, especially in mobile networks, tend towards solutions that would ideally reach the fixed networks' performances, i.e., 1 Gbps everywhere and all the time. Both fix and mobile networks are converging but the technological diversity i.e., Digital Subscriber Line (DSL), fiber, 2G, 3G, 4G, WiFi, tends to make it very complex. A final user could be served by multiple networks, and will choose the best one. This implies the existence of overlapping networks whose interoperability has to be optimized in order to reduce the consumed energy.

Operators are also densifying/updating their networks to increase their perceived Quality of Service (QoS) to such an extent that today we speak of capillary networks or networks with multiple antennas serving an active user i.e., more antennas than active users in a given moment. This would definitely make green networks a very challenging objective.

Furthermore, note that the mobile handsets easy handling, ergonomic and autonomy induce that today 80\% of mobile use occurs in fix situations, i.e., indoor, home, office. Also, the transport of voice for the same quality requires 64 Kbit/s in the legacy Plesiochronous Digital Hierarchy (PDH) network and 320 Kbit/s when it is carried by the IP.

We could think that the Shannon's law is a fatality since to improve the throughput or the QoS it is mandatory to increase the power and therefore the consumed energy. However, other degrees of freedom exist, e.g., the use of smaller and lower power transmitters, the use of Multiple-Input Multiple-Output (MIMO) which allows going beyond Shannon's law by multiplying the coherent transmitters, spatial-time energy focalization, etc. All these techniques can improve the received power without increasing the transmitted one. In this document we aim to give some ideas in this direction. \\

\subsection{Moore's Law then Koomey's: A hardware advance counteracted by software}

Moore's Law is about the computing power of microprocessors. The first Moore's Law that was heuristically established was published in 1965. It stated that the semiconductors’ complexity doubles every two years at a constant cost. The revision of this law in 1975 showed that the density of transistors on microprocessors on a silicon chip doubles every two years. 

This implies that the capacity of the different processors in our servers, DCs or signal processing devices doubles every two years with a chip of the same size. This has greatly miniaturized the Printed Circuit Boards (PCB) for telecommunication equipments such as PCs, phones or tablets.

Dr. J. Koomay resumed this work in order to analyze the energy performance of microprocessors. He published a thesis where he showed that the number of calculations per energy unit consumed is doubling every two years since 1946. We can then think that to reduce energy consumption of telecommunication systems, we just need to renew equipment’s at intervals calculated according to the Return on Investment (ROI). In fact, this evolution is largely absorbed by:
\begin{itemize}
\item The exponential traffic increase and thus the number of operations to be performed per time unit.
\item The physical limits of miniaturization. 
\item The software complexity increase becoming more and more resources greedy. See for example the evolution of the Windows® system through the last 10 years compared to PC capabilities.
\end{itemize}

The main result was that the EE, i.e., number of computations per kWh, of electronics is continuing to grow at the same speed as Moore's law, however, the crucial question is: is there a limit? We have probably to seek a response in the next years from nanotechnology progress. 


\section{Future trends in semiconductors for ICTs}
\label{section2}
In our networks, electronic components constitute an important part in the overall energy consumption. Electronics strongly impacts the performance and the energy consumption of servers, routers, access points and even the end of the optical connections. Also, the energy consumption introduced by electronics in mobile and Internet of Things (IoT) networks has to be taken into account.
 
\subsection{Trends in materials for ICT}

Electronic components can be split into two kinds according to their functionalities. These two classes are defined by the semiconductor used. 

- Silicon components:  Silicon is the most widely used semiconductor. The majority of common components are made with silicon e.g., transistors, capacitors, some photodiodes, etc.

- III-V components: These semiconductors are made with III-V semiconductors as Gallium Arsenide (GaAs) or indium phosphide (InP). They are used for lasers, diodes, Infra-Red photo detectors or some high speed electronic components such as RF transistors.

This separation between silicon and III-V components is present in semiconductor industries. Silicon industries are represented by very huge firms such as Intel, AMD, STMicroelectronics, and Motorola. III-V semiconductors industries are represented by more specialized firms such as Alcatel Lucent for Telecoms, Osram and Phillips for lighting applications, Freescale for high power transistors and Hamamatsu for optoelectronics. In the future this segmentation will be different with the Gallium Nitride (GaN) semiconductor use rise. The boundary between electronic devices and optic devices will probably be less clear than nowadays.

For 10 years now a new III-V semiconductor is emerging for optoelectronics applications as white leds, laser, blue-ray player and high power applications such BSs power amplifiers. These families of semiconductors are GaN and compound of Indium Gallium Nitride (InGaN).

Looking at the evolution of the integration of transistors over the past 75 years, we notice that they have evolved from 1D transistor towards integrated circuits (2D) and nowadays to 3D integrated circuits. This `S' evolution has experienced remarkable and important technological advances that have allowed reducing the consumption and has facilitated the integration/manufacturing.

However, this development tends to slow down as the miniaturization efforts in nanotechnology face significant challenges. This should induce, unless there is a paradigm shift, an increase in the consumption proportional to the processing speed. Some studies already show that we are in the slowdown period for Complementary Metal-Oxide Semiconductor (CMOS) technology processors. The micro-electronics presents a deviation from Moore's Law in recent years, induced by the extreme miniaturization which faces the problem of thermal effect.

It is possible to use low-power design techniques at different stages of a system design. The top-down design goes from the most abstract level to the lowest level. For systems, there are four levels: 

\begin{itemize}
\item Functional level
\item Architectural level
\item Logic level
\item Electrical and physical level
\end{itemize}

Clearly, it is more interesting to implement low power conceptions early in the design process than at the end. At all levels, the method consists on estimating the consumption of the different solutions that are available to choose the best. Current research focuses on the estimation algorithms. It should be considered that the optimal solution for consumption may be disadvantageous from the viewpoint of the processing speed and circuit size and therefore its price. In the future, nanotechnologies and new materials will be more and more present in semiconductors devices and physical limits will be reached.

The solution comes therefore from a reorientation of research on other levers than miniaturization, such as massively parallel architecture. The issue that arises is to delegate some calculations to full optical components that are particularly suited, rather than the massively parallel architecture. Thus, the optical technology will move from purely transmission functionalities to data processing or computing features. This is a highly complex area since the optical components are both expensive and difficult to adapt to PCBs. 

\subsection{Future trends on components} 

The two components that are more strongly impacted by the evolutions of semiconductors are processors and the high power amplifier for BSs.
IoT needs cheap processors with good performance and low power consumption. These characteristics have been implemented in smartphones and now are used in small devices like tiny computers Raspberry or Arietta. The performance has been increased by a factor of three, the power consumed has been divided by three and the cost reduced in a factor of five. These trends are determinant for the future evolution of silicon industries since they give a positive balance between smartphones’ mobility and performances.

However, silicon has limited physical properties. Semiconductors such as GaN allow overtaking these limits. GaN components will work at higher voltage i.e., more than 1kV, several A/mm and at high frequency level i.e., more than 30 GHZ. 


GaN can be produced at a lower cost than silicon and it enhances the transistor performance. With GaN transistors, there is an ability to switch higher voltages and higher currents faster than any other transistor. With these advanced characteristics, new applications can be introduced. 

Industries control already the growth of GaN on 150 mm silicon wafers and start tailoring the growth on 200 mm silicon wafers. They managed to replace the sapphire, which is much more expensive than silicon. All these headways show that the GaN components should quickly replace silicon, particularly for RF transistors used for power amplifiers in BSs.


\section{Optical networks}
\label{section3}

The network transformation to all-optical is obviously a sustainable and effective way of reducing the energy consumption. It primarily affects the transport and routing in the network sector. Its advantages are: 
\begin{itemize}
\item No losses in transmission.
\item Immunity against electromagnetic interference.
\item A significant bandwidth.
\end{itemize}

Today, it is estimated that an optical fiber can carry up to 160 wavelengths with the Wavelength Division Multiplexing (WDM) technique. Each wavelength can support 10, 40 or 100 Gbit/s and soon should achieve the Tbit/s. Notice that due to Shannon's limit, the maximum capacity will be around 50 Tbit/s and therefore the number of wavelengths will be less than the 80 transmitted today.
Currently, these high capacity optical fibers are primarily used to provide point-to-point interconnections between electronic routers. In the following sections, we address the envisaged solutions in each network sector to make full-optical networks more energy efficient.

\subsection{Core and transport} 
The data rates evolution from 10 Gbit/s nodes to 100 Gbit/s nodes has increased the capacity of the optical fiber links. However, it did not imply power saving since the new transponders consume more. For example, a 10 Gbit/s transponder consumes around 30 W, and a 100 Gbit/s transponder consumes around 160 W i.e., an increase of 10 times the data rate and 5 times the energy consumption. The envisaged solutions are the following: \\

\subsubsection{Bypassing nodes} 
In optical networks, the important points are the capacity and the reach. Indeed, the reach allows saving in the number of Optical/Electrical (O/E) conversions. Having long reach elements allow to bypass nodes and therefore to reduce the O/E conversions. This bypass is often called ``transparent offloaded". Transparency in the core networks avoids the need of signal regeneration, the use of routers, and all the ``useless'' conversions. This technique showed that 50\% of energy reduction at the core and metro network can be achieved. It has been introduced in Orange networks since 2008. The motivation to do so was economic rather than saving on network energy consumption \cite{ref5}.

The idea is not to remove the electronics of the transport layer because there will always remain converters, but to make the most of optics. The objective is to reduce the electrical conversion needed for the transit or to better share resources to reduce their number. \\

\subsubsection{Photonic integration} 
This is a very promising point which integrates the signal processing and the laser on the same silicon board (hybrid photonic-electronic chip). This innovation will considerably reduce the consumption of the E/O conversion and thus an improved end-to-end connection.
 
This lever is currently being integrated in equipment by ALU and Huawei. Nevertheless, it is in its very first stage and we believe that further integration will enable more energy savings in the coming years. 

Summarizing, this technology is currently being applied punctually for certain long distance optical transmission equipment in the core network. In the coming 3 to 5 years it is expected a generalized application at least in all the O/E conversion parts of all equipment with optical interfaces. For the 5-10 years to come, we could expect to have this technology in the DCs for the servers’ interconnection with a massive integration in order to decrease the cost and complexity \cite{ref6,ref7}.

Finally, operators will not participate in the development of this technology. Nevertheless, increasing our requirements in terms of cost and consumption will push vendors to boost this solution or maybe others.

\subsubsection{Dynamic routing} 
The network is a mesh where multiple paths between two nodes are possible. In the one hand, the transport can be made via a “single path” which may seem more economical. The problem is that this path must be oversized and cannot be disabled. On the other hand, the use of diversity of routing paths allows transmitting the information through different routes and to turn off or pause the paths without traffic based on the load.
Adaptive rerouting of traffic associated with Low Power Mode can also bring savings thanks to the large traffic variation observed during night and day periods. This is particularly profitable if multiple paths between nodes are possible. Indeed this provides the finest granularity to switch off capacity according to effective load, and this may also reduce the impact on latency variation. Despite the fact that some technology enablers exist, the overall impact on the QoS or network reliability is still unknown and studies are needed to prove the applicability of adaptive rerouting.

\subsubsection{Towards full-optical switches} 
The Optical/Electrical/Optical (O/E/O) conversion can be reduced and eventually eliminated through the optical switching that would allow data to be switched directly in the optical domain. The easiest way to provide connectivity in an optical network is by assigning a specific wavelength to each source-destination, this technique is called Optical Circuit Switching (OCS). By bypassing the intermediate nodes in the optical domain, the OCS is able to eliminate the need for O/E/O conversion in the core of the network. However, in the OCS, the whole capacity of a wavelength is dedicated to a specific source-destination pair and cannot be shared by other nodes. This coarse granularity in the switching can lead to severe bandwidth under-utilization, since one wavelength can carry up to 100 Gbit/s.
Today, electrical solutions are used to exploit correctly the coarse granularity. Tomorrow, photonic switching solutions could also address this issue providing both capacity and fine granularity thanks to optical packet or burst switching. Studies \cite{ref8} have estimated that savings could be very interesting i.e., more than 20\% in metro networks depending on scenarios and traffic. Such technologies are thus actively explored notably in Orange Labs, and there have been few industrial prototypes from ALU, INTUNE, Huawei, among others. Industry and operators seem however chilly to evolve towards these solutions, making technology progress slower than we would like. From our point of view, this lever is feasible, however, it still lacks of maturity. \\

\subsubsection{Removing redundancy} 
In optical networks and particularly in the core, redundancy is a simple and effective way to ensure the continuity of the service. This implies that every element of the network is duplicated (1+1) and permanently active, with the implicit extra energy consumption. Furthermore, it requires bandwidth reserves that would be used only punctually.

Some ideas are beginning to emerge under the name of Quality of Protection (QoP). The principle of this technique is to define a QoP indicator as a function of the service and to reduce the QoS for lower priority services. The QoP is based on an optimization strategy of the transmission paths according to the required flow rate and the energy consumption. 

The expected energy savings could be quite interesting i.e., several tens per cent depending on the scenario. However, this would require a deep change in “mentality” of operational teams since it means trading network availability against energy. Any disruption in the core transport network, where protection is implemented, can impact hundreds of thousands of customers. Giving up on protection for low QoP traffic means that for these clients there could be large interruption of service in case of a failure.

It is worth mentioning that this lever is very sensitive for an operator as it directly impacts the security and the service reliability. This would need a complete paradigm shift on network security applying other technics than just redundancy. From our point of view, this lever is far from being implemented. \\
\subsubsection{Access} 
Orange has experienced full-optical networks as a first step towards switching to an all-optical access. This will make obsolete superposed networks dedicated to a single service i.e., Public Switched Telephone Network (PSTN), DSL, residential. The performance in terms of capacity and range will conditionally allow:
\begin{itemize}
\item The unification in all-optical from central office will allow decommissioning the copper lines. 
\item The use of optical interfaces for residential, businesses, wireless BSs connection and isolated equipment. 
\item The introduction of optical equipment higher in the network infrastructure, which distances typically are up to 20 km. 
\item The use of amplification or regeneration techniques, known as reach extender, will also allow reducing operating complexity and the required consumption of the small central offices.
\end{itemize}

In the access, additional energy saving opportunities will result from the implementation and activation of standby modes and/or by the capacity dynamic adaptation as a function of the instantaneous capacity required by the service. For point-to-point communications, it is imperative to develop an optical transposition of the IEEE 802.3az \cite{ref9} standard dedicated to the Gigabit Ethernet copper interface.

In current optical networks deployments (``LA Fiber" Orange project) sleep modes exist for all G-PON equipment as well as for the future XG-PON1 or NG-PON2. However, the dynamic adaptation of the service transmission rate is being introduced using multiple modulations, which will potentially introduce dynamicity to the hardware consumption as a function of the data rate.

Finally, when considering the transition to full-optical networks it is needed to consider the emergence of multiple low rate applications such as Low-Power Wide-Areas networks (LPWA), which were not envisaged in the initial deployments dedicated to triple play residential IP. These terminals will not introduce big amount of data but will prevent the network from entering a sleep mode due to their recurrent transmission. 
%
%
%

\section{Radio networks}
\label{section4}
In recent years the radio industry has experienced a real change from the green point of view. This change has not been the answer to the environmental problems our planet is facing but to two practical reasons: on the one hand the growth on the requirements of autonomy of user devices and on the other hand the technological radio expansion to developing countries with no or limited electrical infrastructure.

In wireless communications the information is transmitted through electromagnetic waves traveling over the air. These waves are subject to the path loss which is the loss in signal strength when traveling from the source to the receiver. Conservation of energy tells us that signal will reduce as a function of the distance to the receiver.

It is very interesting to realize that for 1 W transmitted by a wireless Access Point (AP), the power perceived by the User Equipment (UE) is not more that 10$^{-6}$ W, which means that 99,9999\% of the signal is lost elsewhere. It is astonishing to find that the energy waste is fundamentally inherent to the wireless cellular system. The energy loss is therefore in the DNA of wireless systems, fact that should be hopefully addressed for 5G. 
\subsection{A 5G designed to be green} 

The next generation mobile communications, or 5G, embraces ambitious objectives. It is expected to provide 1000 times more capacity than 4G, to support 9 billion users and a more diverse range of applications, services and device types. To achieve these targets, 5G should reduce latency, improve the communications reliability, have longer battery life for devices and higher user bit rates. All this with a drastic enhanced EE which will enable 5G systems to consume a fraction of the energy that 4G mobile networks consumes today for the same transmitted amount of data. From white papers published by consortiums, manufacturers and operators we see that EE has been put forward as one of the main targets for 5G. Nevertheless, few details on how this objective could be reached are given.

Figures on green have been given by Asian and European companies. On the one hand, some Asian companies bet for a 5G network that will keep the EE stable with regards to 4G (Datang). On the other hand, some European companies envisage a greener mobile network where the consumed energy is divided by 10. This number has been given by EU METIS project and adopted by 5G PPP association.

Orange appears as the only operator that advocates for the introduction of the ambitious green target of reducing the network consumed energy by a factor of 10 \cite{ref10}. 

For the green network aspects, two Key Performance Indicators (KPIs) are indistinctly used: EE (bits/Joule) and energy consumption (Watt). In order to conceive an optimal network in terms of energy consumption, the green requirements have to be considered from the beginning, unlike previous mobile system generations for which energy consumption concerns were addressed on the fly.

Despite the fact that EE is put forward as one of the core characteristics of 5G by all players, especially by Orange Group, the green commitments of some consortiums remain hazy. Firstly, there is not an agreement on the energy KPI to be used and how to measure it. Secondly, figures given by different players are divergent. Last but not least, the embedded energy and the environmental impact of the required material for the implementation of 5G, foreseen at the 2020 horizon, should be considered. In order for 5G to become the revolutionary technology that new social tendencies will impose it is clear that it has to be natively green. 
\subsection{Base stations power model evolution} 
It is well known that the mobile access equipment consumes 80\% of the total mobile network energy \cite{ref11}. From this 80\%, nowadays 90\% is consumed by the BS and the remaining 10\% by the backhaul. In France there exist more than 20.000 sites, most of them equipped with multiple technologies. Reducing or optimizing the BSs energy consumption is therefore essential to achieve real energy savings.

\begin{figure} [ht]
\centering
\includegraphics[width=8.5cm]{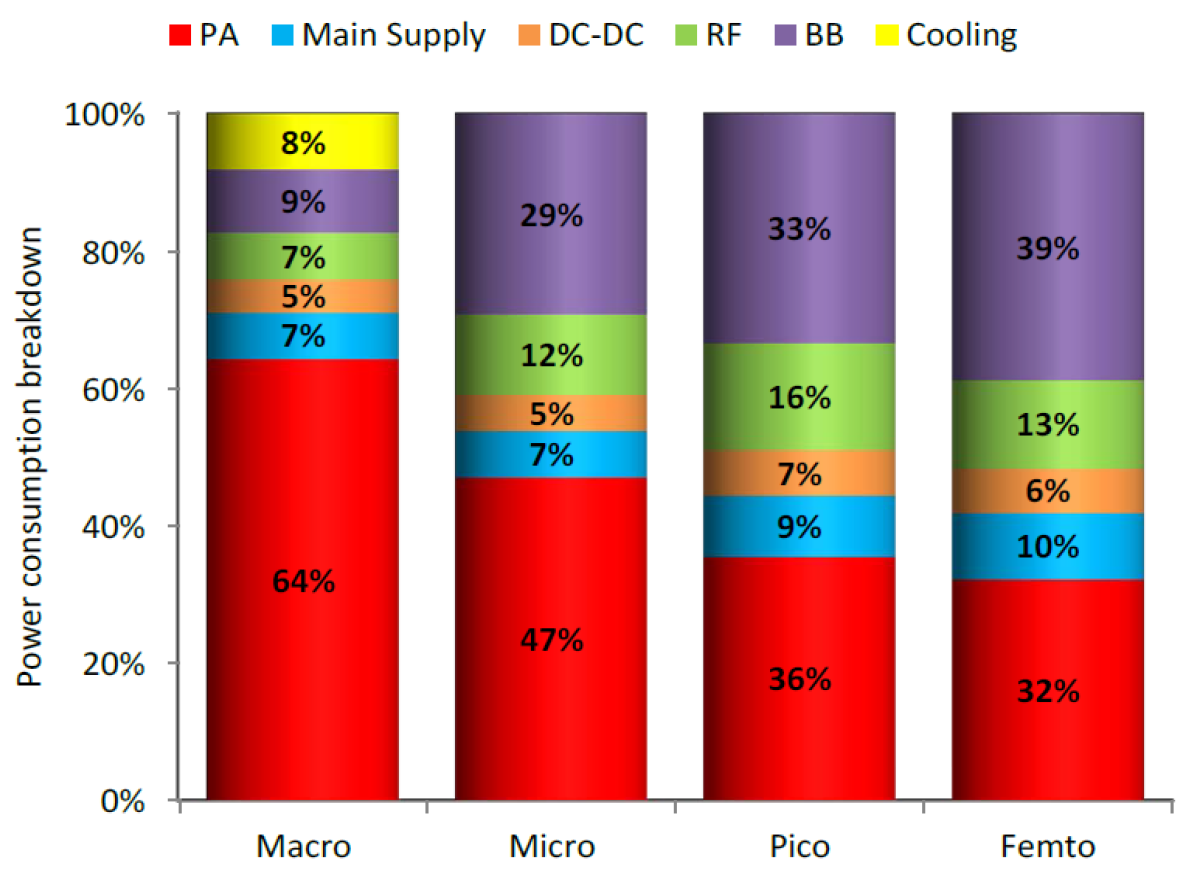}
\caption {Power consumption breakdown in different types of BSs} 
\label {PC_Breakdown}
 \end{figure}

Figure \ref{PC_Breakdown} presents the breakdown of the energy consumed by the different parts of each type of BS. This is a guide for researchers to show them were efforts are needed. For instance, in macro BSs the effort should focus on the Power Amplifier (PA) optimization, while for small cells, the Baseband Unit (BBU) and PAs are equally significant. Anyhow, many research groups are working on this topic and interesting advances are being achieved, notably, sleep mode techniques are very promising.

Research studies have shown that the BS power consumption is composed of two parts: 1) architectural costs ($P_0$): fixed processes such as control signaling, backhaul infrastructure, and the load-independent consumption of baseband processors which does not scale with the load and 2) transmission processes: such as transceiver chains, coding and decoding and channel estimation and precoding with an energy consumption that linearly vary with the traffic.

The power model for next generation networks should evolve in such a way that with the integration of new techniques and more power proportional components (i.e., GaN technology for PAs, envelope tracking, etc.) and systems, we should arrive to almost zero consumption at zero load. Then, with the integration of more energy efficient components and processes, the energy consumption at maximum load should be decreased.

Nowadays research focus on decreasing the power consumption at minimal load, $P_0$, which is the straightforward way to make mobile networks greener. Notably, advances in hardware during the last years have allowed shifting $P_0$ from 80\% of the total BS consumption to less than 50\%, for the newest technology. Thanks to that, current power models have 50\% fix consumption and 50\% that scales with the load. 

The existing sleep modes are very basic since their only target is to not degrade the QoS and to remain transparent to the communication protocols. This results from the fact that green concepts became important after the air interface design. Nevertheless, recent research \cite{ref12, ref13} show that important gains in energy savings could be achieved with the design of more sophisticated and comprehensive sleep modes, e.g., deep sleep mode, which would require a change on the air interface i.e., duty cycle, frame structure, etc. this will be possible with the new air interface design in 5G. 

\subsection{Advanced sleep modes}
\label{ASM}
Nowadays, only fast sleep mode is proposed and used by operators. It was quickly adopted by operators due to its short transition time, i.e., symbol rate, which therefore does not affect the users’ QoS. This type of sleep mode brings an energy consumption decrease of about 10\% at the telecom part of the BS which represents around 3\% at the site level.

Nevertheless, as presented in the previous section, new telecommunication systems should have almost zero consumption at zero loads. This can be achieved with ASMs, which consist on putting systems in different sleep modes levels depending on the idle periods. However, ASMs will imply the introduction of a new frame filling including signaling, which can be challenging.

The BS power model developed by GreenTouch comprises multifold hardware components. This model introduces the ASM feature where each of the hardware components can be disabled or configured in a power saving mode when not used. The model introduces four sleep levels that combine hardware sleep modes depending on the idle periods. 

The BS is considered to work in different sleep modes depending on its load. The operational modes are: 
\begin{itemize}
\item Sleep mode 1 (SM$_1$): The power amplifier and some processing components are disabled on an OFDM-symbol level (micro sleep). This can be used at reference symbol level. It takes 71 $\mu$s to deactivate and to activate again (transition time). With this sleep mode it can be reached around 15\% of energy savings. 
\item Sleep mode 2 (SM$_2$): For this sleep mode more of the processing hardware is deactivated, so a lower power level is reached. However, the transition time is estimated to 1 ms i.e., subframe level. With this sleep mode around 35\% of energy savings can be achieved. 
\item Sleep mode 3 (SM$_3$): This is a yet deeper sleep mode with more components switched off and the transition time for both directions is 10 ms, i.e., interrupts the transmission for at least one radio frame. 
\item Sleep mode 4 (SM$_4$): This is the standby mode where the BS is out of operation but retains wake-up functionality. Additionally the backhaul is active to re-activate the BS. The transition time is more than 1 s . 90\% of energy savings can be achieved with this sleep mode. For instance, if a new air interface is designed in such a way that it has more than 1 second between two consecutive transmissions, then the SM$_4$ can be applied.
\end{itemize}

Table \ref{tab:paramsg} summarizes the characteristics of the different ASM levels: 

    \begin{table}[ht]
    \small
    \renewcommand{\arraystretch}{1.3}
    \caption{Advanced Sleep Modes characteristics}
    \label{tab:paramsg}
    \centering
    \begin{tabular}{|c|c|c|c|}
    \hline
    \textbf{Sleep} & \textbf{Deactivation} & \textbf{Minimum} & \textbf{Activation} \\
     \textbf{level} & \textbf{duration} & \textbf{sleep duration} & \textbf{duration} \\
         \hline
    \textbf{SM$_1$} & 35.5 $\mu$s & 71 $\mu$s & 35.5 $\mu$s \\
	\hline
	    \textbf{SM$_2$} & 0.5 ms & 1 ms & 0.5 ms \\
	\hline
	    \textbf{SM$_3$} & 5 ms & 10 ms & 5 ms \\
	\hline
	    \textbf{SM$_4$} & 0.5 s & 1 s & 0.5 s \\
	\hline 
    \end{tabular}
    \end{table} 

Orange is collaborating with manufacturers in order to boost the introduction of ASMs since they are very promising and align with optimal telecommunication operational mode, the almost zero consumption at zero load. The arrival of 5G technology appears as the perfect framework for the introduction of this green technique. \\

\begin{figure} [ht]
\centering
\includegraphics[width=8.5cm]{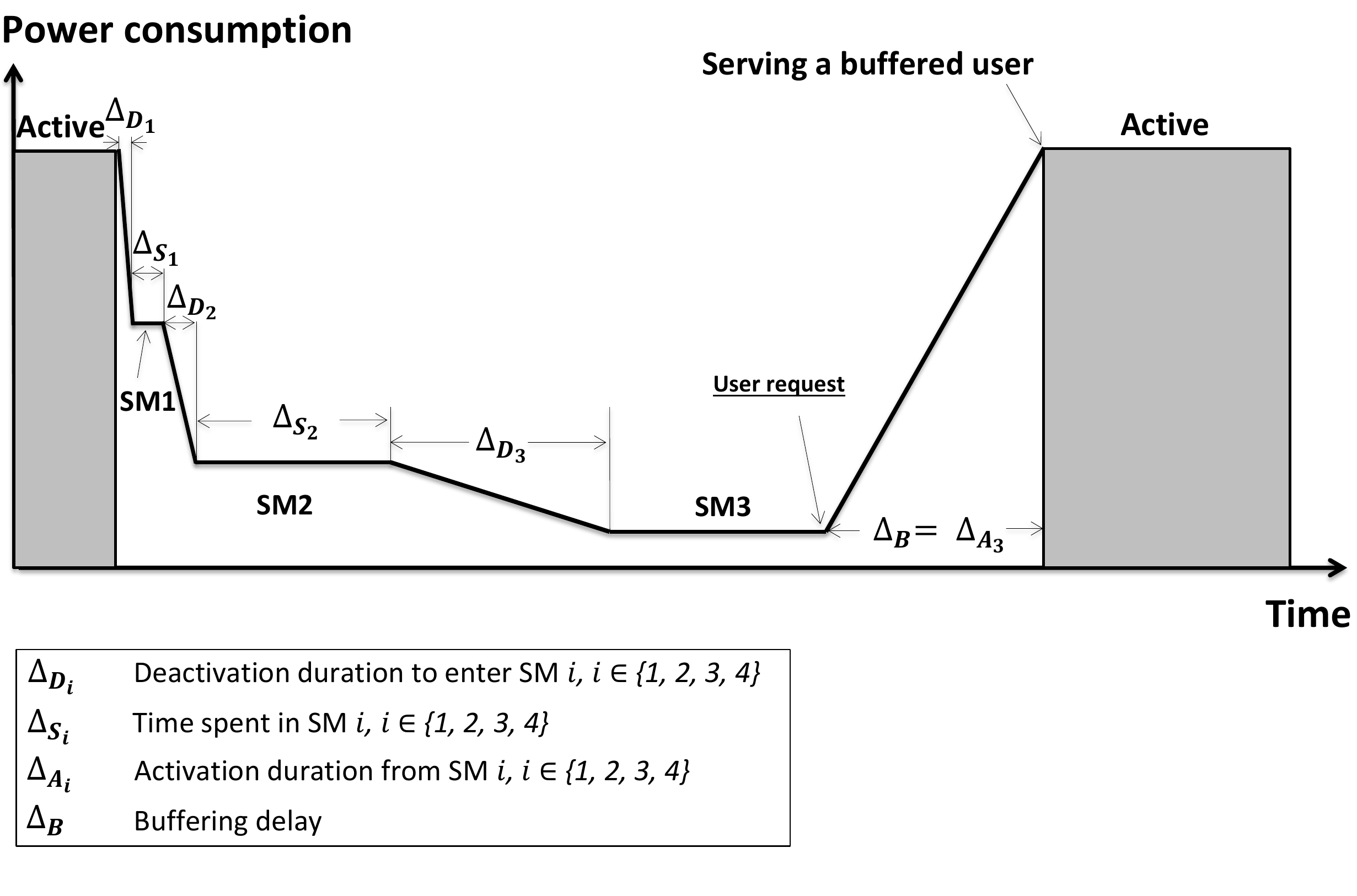}
\caption {Example of implementation strategy of ASMs} 
\label {fig:strat}
\end{figure}

\textbf{Example of implementation strategy} \\
A simple way to use the ASMs is a gradual deactivation of the BS, i.e., whenever the BS is idle (not serving any user), we can put it into the different levels of sleep modes gradually going from the shortest one to the deepest as shows Figure \ref{fig:strat}.

If a user requests a service, such as download of a file, while the BS is in sleep mode, we put it in a buffer and trigger the activation of the BS. The considered user has then to wait until the BS wakes up and becomes ready to serve him. Therefore, we can be confronted to latency increase due to these buffering durations.  

\begin{figure} [ht]
\centering
\includegraphics[width=7cm]{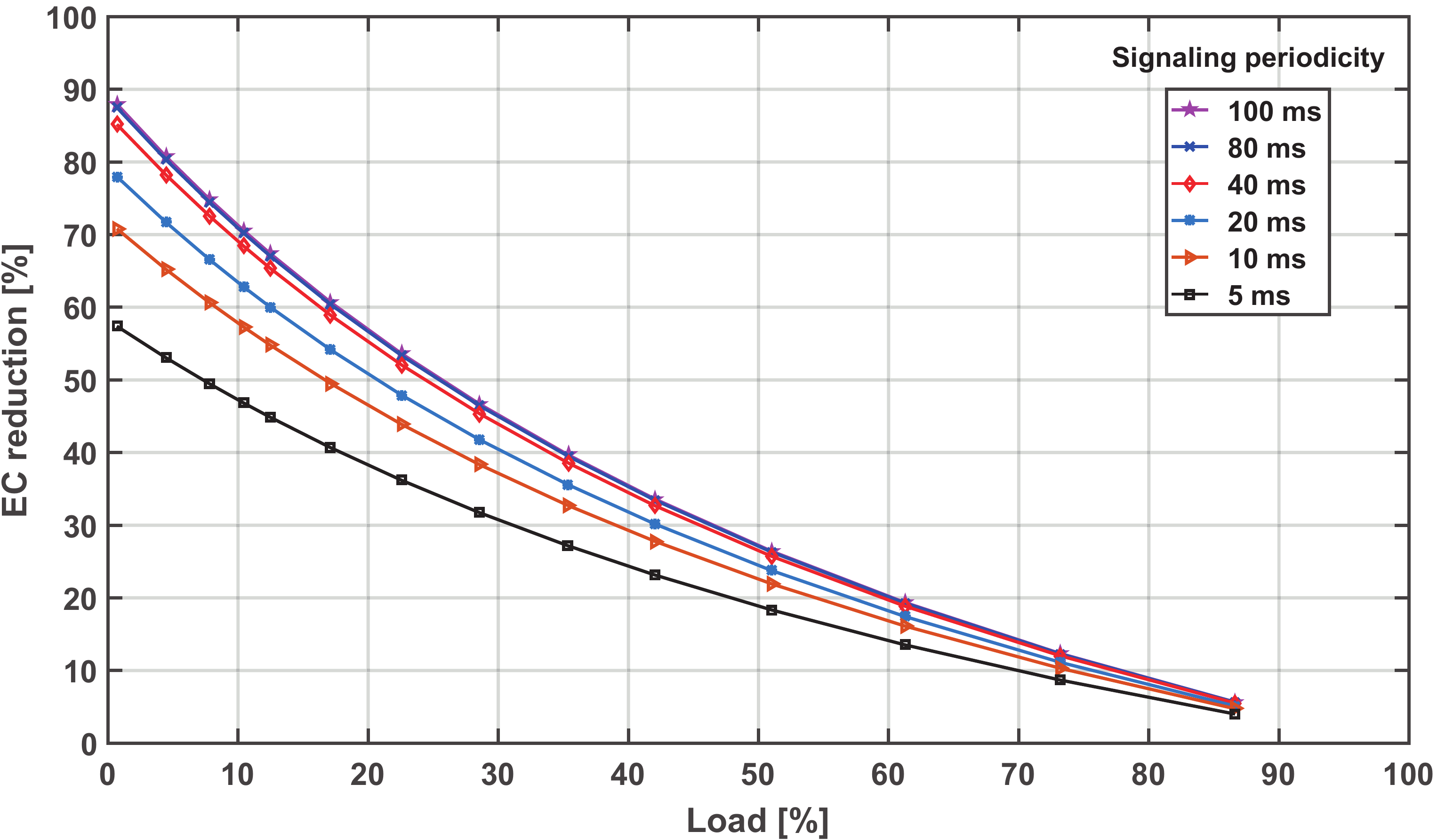}
\caption {Energy consumption reduction using ASMs depending on the signaling periodicity}
\label {fig:EC}
\end{figure}

\begin{figure} [ht]
\centering
\includegraphics[width= 7 cm]{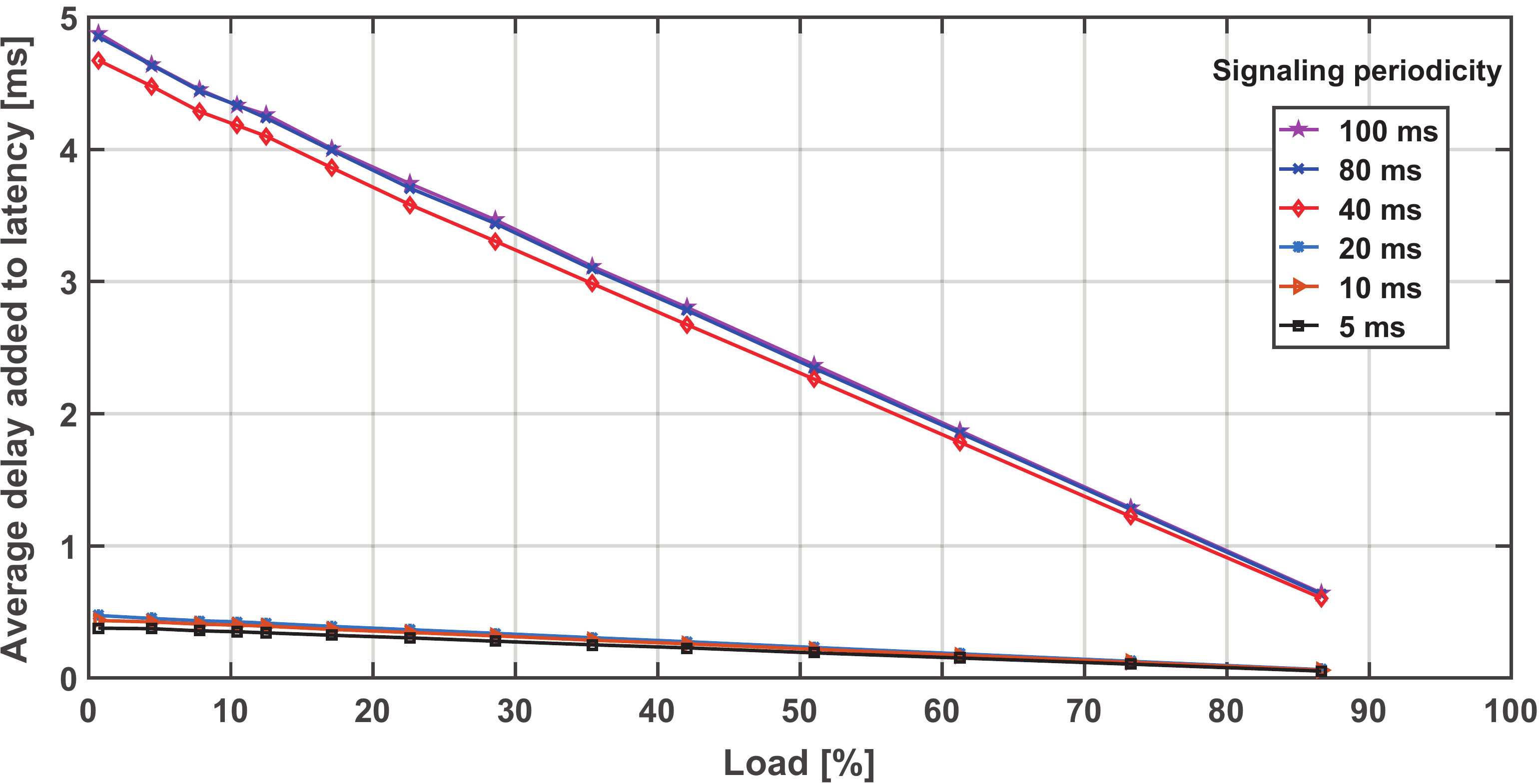}
\caption {Average delay added to the baseline latency when using the ASM}
\label {fig:latency}
\end{figure}

The Bs has also to wake up periodically in order to send synchronization bursts such as Primary and Secondary Synchronization Signals (respectively PSS and SSS), also Physical Control CHannel (PBCH). With such frequent synchronization signals, we cannot use deep sleep modes, so that we have less energy reduction. We studied in \cite{ASM} the impact of an increased periodicity of signaling both on energy consumption and network performances. Figure \ref{fig:EC} shows that up to 90\% of energy savings can be achieved in low loads when we increase the signaling periodicity. This energy reduction is decreasing with the load since the more the load increases, the less time between consecutive users is available to put the BS into sleep mode. 
However, this strategy induces a latency increase which can achieve 5ms as shown in Figure \ref{fig:latency}. This is a constraint in some critical scenarios like Ultra-Reliable Low Latency Communications (URLLC) use case in 5G networks which requires a latency less than 1ms. Therefore, a smart orchestration of the ASMs is needed according to the requirements of the network operator in the different use cases and scenarios.  	

\subsection{Amplifiers} 
The PA market is dominated by the freescale society \cite{ref14}. Nevertheless, there is an increase in the number of manufacturers working in this field such as Huawei, NSN and ALU. Their objective is to differentiate from other manufacturers by conceiving and producing their own chips and to develop their own strategy. Indeed, some manufacturers are moving towards more power per band amplifiers (European manufacturers) while others favor the multi-band amplifiers (China). We consider that per band amplifiers optimization is a better approach since nowadays, there is no technology mature enough to achieve a good efficiency for multiband power amplifiers.

The development of Radio Frequency Power Amplifier (RFPA) touches on different areas such as electronics, materials and signal processing. The three drivers in the amplifier development are the efficiency, instantaneous band and linearity.

The PA represents 70\% of the consumption of the telecommunications parts of a BS. Note that the PAs are constantly improved but the changes in protocols and technology, i.e., passing from TDMA to CDMA to OFDMA, often challenge these improvements, especially with the increasing need for bandwidth and modulation.

The main role of a PA is to amplify a signal as closely as possible to the original one without introducing distortion. There are two major classes of RFPA: 
\begin{itemize}
\item The ones used for non-linear signals amplification (class C) as for GSM.
\item The ones used for the amplitude modulation signals (class AB) such as CDMA and OFDM, which require linearity capabilities.
\end{itemize}
 
In the first case, the RFPAs are simple and they have very high efficiencies. The performance in terms of EE of 3G PAs have increased from 15\% to 45\% in 15 years. For the 4G, PAs still have an EE of 20\%.

Until today, amplifiers are optimized to the maximum power and at 100\% of load. As an initiative of Orange, an ETSI document has been standardized so that the amplifiers efficiency is calculated at several load levels in order to reflect better the actual operation of these electronic components.

Several techniques have been studied to improve the linearity of the PA, but only two architectures have been a commercial success, the Doherty and the envelope tracking.

\subsubsection{Doherty or enveloppe tracking}
Doherty is a clever combination of two known amplifiers, the Class AB and Class C put in parallel and combined. A Doherty behaves strictly as a class AB at low power ($<$10 W). This technology is widely used in the RFPA at the BS.

The envelope tracking allows adjusting the DC voltage to operate the PA at its optimal efficiency. This highly efficient technology is already implemented in the low-power mobile devices such as the terminals. Its extension to high powers remains to be reliable. Also, in the framework of the OPERANET II European project, a proof of concept has been demonstrated until 40 MHz bandwidth. Manufacturers are still working on bandwidth extension to more than 100 GHz.

\subsubsection{The gear stick}
A PA optimized to the maximum power is like a moving car with a gearbox blocked on the last speed. The consumption/transmission ratio is good only for the high powers and it degrades very quickly at low power. This is what is now installed on the mobile networks. Various techniques begin to appear in order to introduce different and multiple optimization points. Therefore, the amplifier will have the capability to transmit different powers with efficiency close to its maximum at each power level.

This trend is real, even if the performance of these techniques are not entirely satisfactory today due to the intermodulation problem, the speed regime change problem, etc. these barriers will probably be overcome in the following 5 years.

\subsubsection{Parallelization}
Based on the MIMO principle that replaces a link between a transmitter and a receiver by several orthogonal links and on the fact that a low PA is more effective than one at high power, the parallelization can be very useful to significantly increase the effectiveness of the BS.

Today, this idea is considered by almost all the manufacturers’ research departments. It has been shown by Huawei, ALU, and Ericson that by combining the reduction in size and the multiplicity can improve the link budget. On the other hand, the Interuniversity Microelectronics Centre (IMEC, ZTE) have shown that this can improve the energy balance. The realization of a prototype is already underway in China by ZTE and CMCC and it is already in test phase at ZTE.

\subsubsection{Focusing antennas} 
Focusing antennas spatially focus the radiated power to the user, in an adaptive way and independently of its position. In comparison with conventional antennas covering a cell, these ``smarter" antennas reduce the required energy to satisfy a user with a gain increased in several dB. Notice that when gain is increased the power can be decreased the same amount of dB.

Some focusing techniques make the received signal both stronger and ``clean" i.e., without echoes. This would allow the use of very simple receivers as those used for 2G but achieving data rates of 4G \cite{ref15}. Even better, focusing antennas are usually composed of antenna elements, each one of them integrating their own amplifier. Research studies performed by ALU and Huawei, show that we can replace a large amplifier inefficient in terms of power consumption by highly efficient small low power amplifiers.

These antennas are used for point-to-point connections, as well as for microwave transmissions. Nevertheless, their use for moving targets is still far to be deployed due to the following reasons:
\begin{itemize}
\item The frequency needed to update the beam increases with the target speed. The standards limit this frequency to around 1 kHz. At most, a pedestrian speed could be supported. However, recent studies show that with the disruptive concept of predictive antenna, it could be possible to maintain a robust beam for vehicular speeds i.e., up to 300 Km/hour \cite{ref16, ref17}.
\item It is needed the synchronization of a large number of antenna elements and radio channels with a sub-nanosecond precision. The electronics for signal processing and the RF cards are rapidly evolving in order to support a growing number of elements i.e., 64 elements nowadays and hundreds in three years’ time, see Figure \ref{mMIMO}.

\begin{figure} [ht]
\centering
\includegraphics[width=5cm]{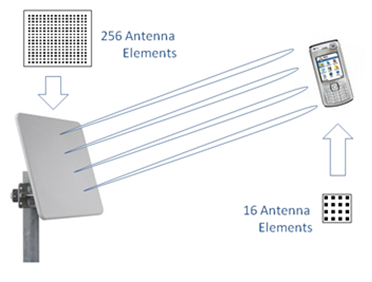}
\caption {Massive MIMO boosts space-time docusing} 
\label {mMIMO}
 \end{figure} 
 
\item The targeted user shall transmit a training signal on the same frequency carrier as the focusing antennas. However, mobile network standards in some cases use different frequency bands for the BS and the terminal. This problem can be easily solved for some focusing techniques \cite{ref18}.
\end{itemize}
\subsection{Energy-free communicating tags thanks to 5G backscattering}
In 2013, the University of Washington presented a new communication system that enables an RF tag, called ``Ambient Backscatter" \cite{refA}, to send data to an RF reader, without battery and without generating any new RF wave. Instead of spending power to generate an RF wave, the RF tag just backscatters the ambient RF waves generated by TV towers. In this sub-section, we will first present our own prototype of backscatter communication system, designed based on \cite{refA}, and then, we will elaborate on the potential applications for IoT cellular networks operators.

\begin{figure} [ht]
\centering
\includegraphics[width=8cm]{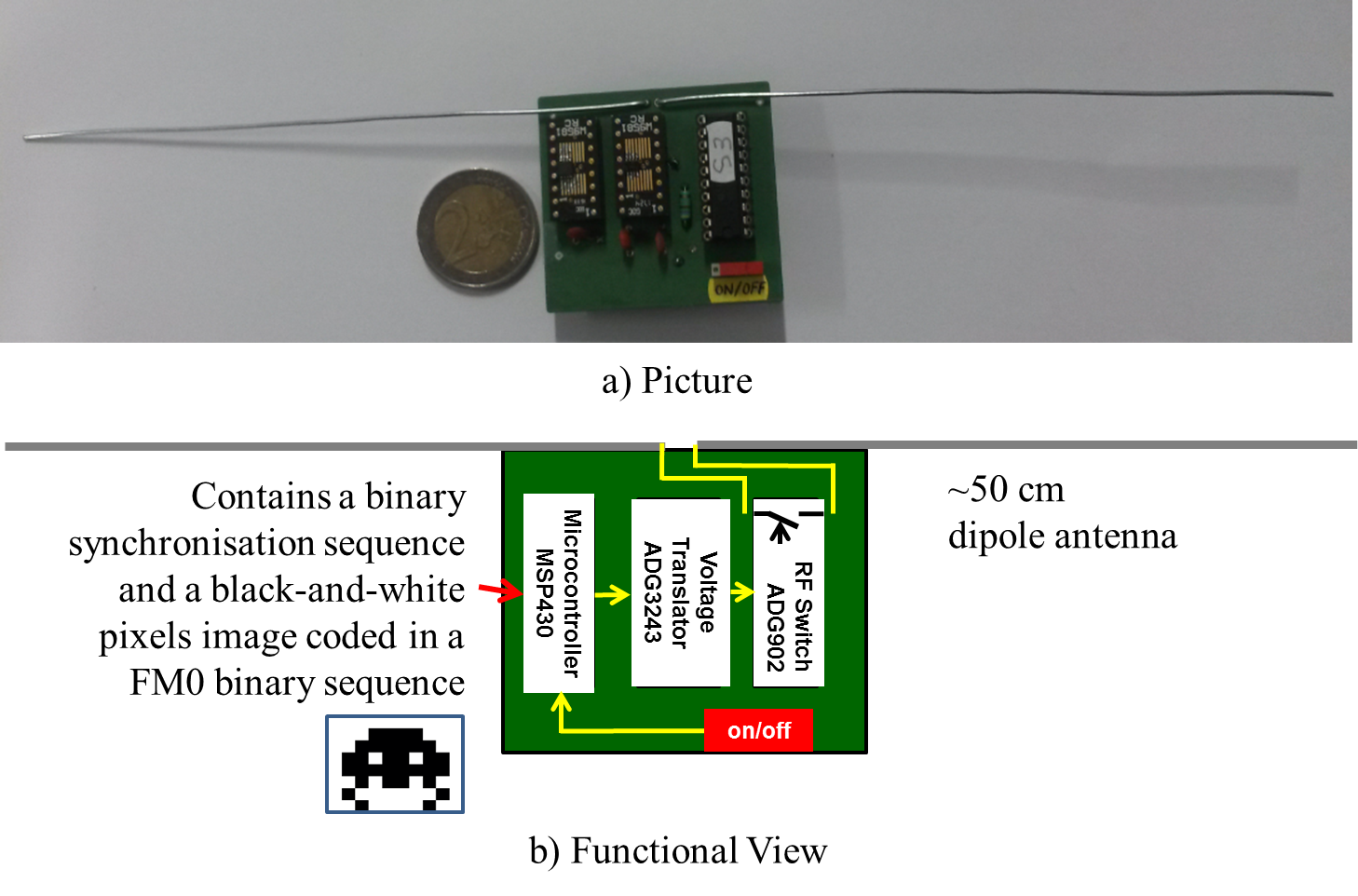}
\caption {Our prototype RF tag designed by Dr. Nissem Selmene} 
\label {RF_prototype}
 \end{figure}

As illustrated in Figure \ref{RF_prototype}, our RF Tag includes a dipole tailored for TV frequencies around 600 MHz, an MSP430 microcontroller, an ADG920 RF switch and an ADG3243 Voltage Translator. The RF switch sets the dipole either to a short-circuit state or an open-circuit state. When the dipole is in one state, it is mainly “transparent” to TV RF waves, whereas when it is in the other state, it is mainly backscattering RF waves. The switch is controlled by the MSP430. This latter component has a FM0 modulation binary sequence stored in its memory (coding for a black-and-white pixels image) and a synchronization sequence. The MSP430 sets the dipole to one state, to send a “0” and to the other state, to send a “1”. As in one state the tag is transparent to TV waves whereas in the other state, it is backscattering them, the reader detects a change in the received TV signal. The message is sent periodically with a data rate of 10 bits/seconds. 

\begin{figure} [ht]
\centering
\includegraphics[width=7cm]{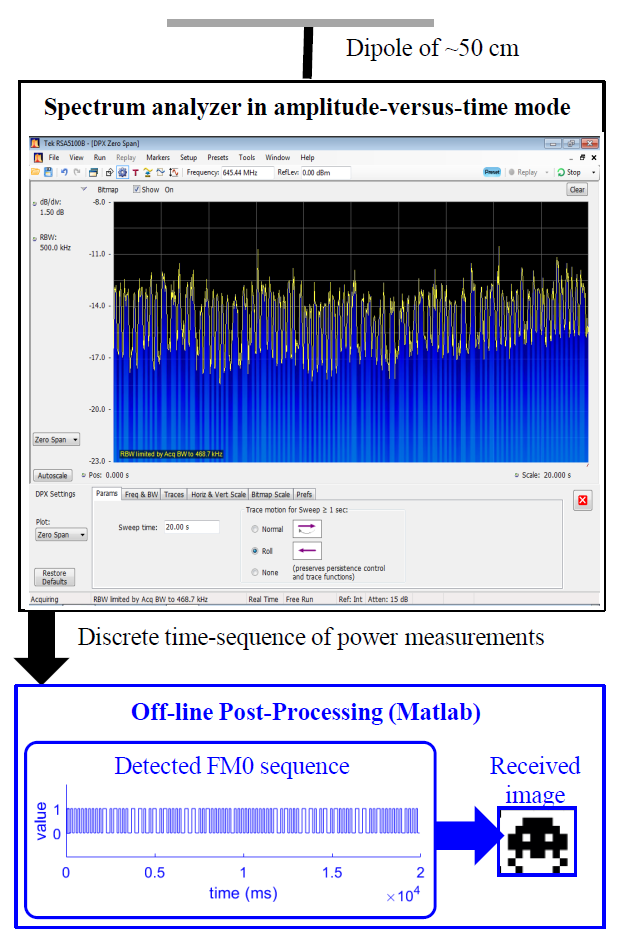}
\caption {Our prototype RF reader} 
\label {RF_reader}
 \end{figure}

As illustrated in Figure \ref{RF_reader}, our TV reader includes the same dipole antenna as the RF tag and a RSA5100B Tektronix spectrum analyzer. The RSA5100B measures in real time, the received power versus time, over a time window of 20 seconds. The power is measured at the frequency of 645 MHz, within a receive bandwidth of 500 kHz.  We store a 20 seconds time-sequence of measurement and then post-process it offline, using a matlab software (SW) code. The SW code simply compares the received power with a moving average power threshold to determine the changes in the received power level. Then, the synchronization and the FM0 demodulation is performed, and the original image is retrieved. 

\begin{figure} [ht]
\centering
\includegraphics[width=8.5cm]{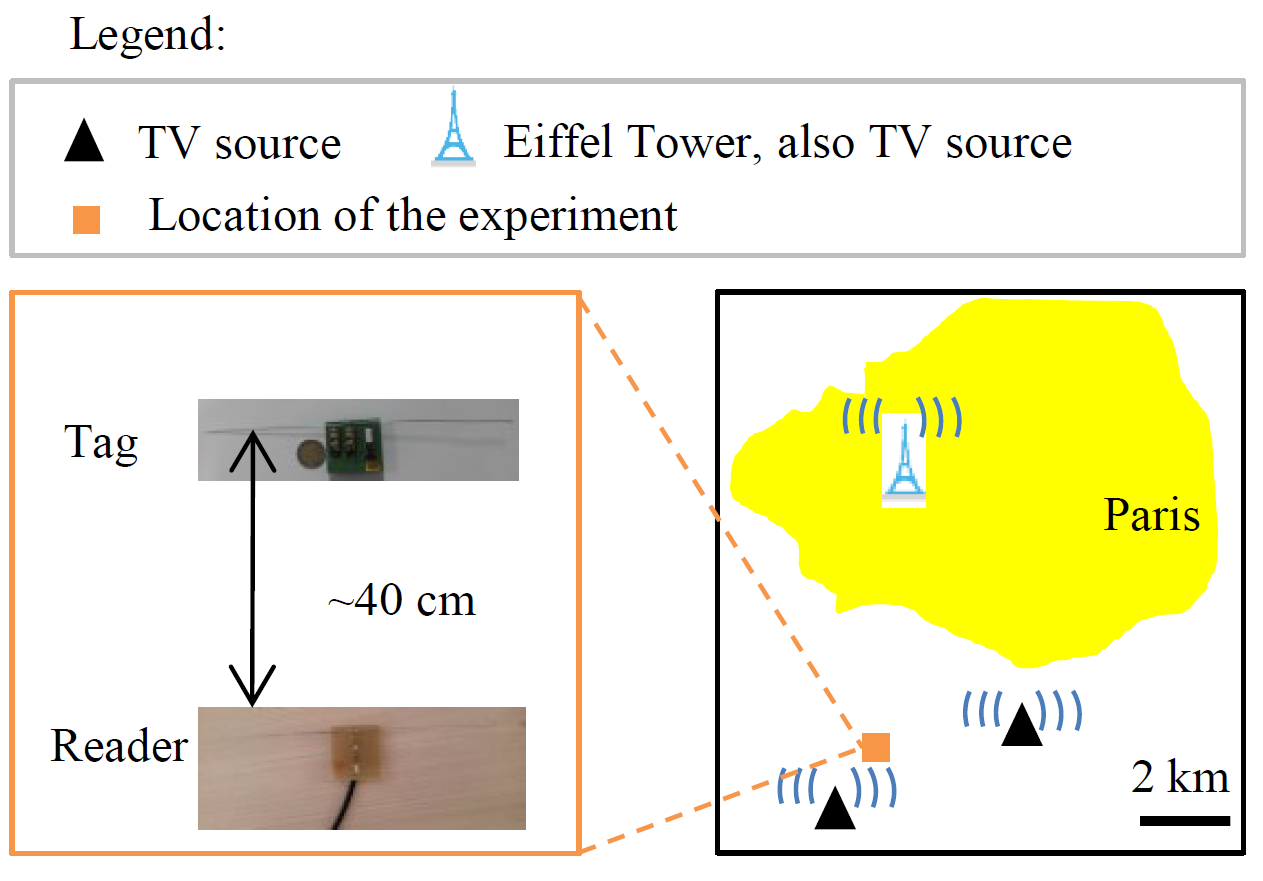}
\caption {Distances between reader, source and tag during the experiment} 
\label {Distances}
 \end{figure}

Figure \ref{RF_reader} illustrates an actual measurement and the corresponding successful demodulation, performed in Orange Gardens, Chatillon. As illustrated in Figure \ref{Distances}, the nearest TV source was 2 km away from our location and the tag-to-reader distance was of around 40 cm (i.e. almost a wavelength). The experiment was performed indoor at the ground floor, near a window. According to works summarized in \cite{refB}, one can expect higher tag-to-reader distances (several meters) with more advanced detection algorithms than a basic energy detector.

\begin{figure} [ht]
\centering
\includegraphics[width=9cm]{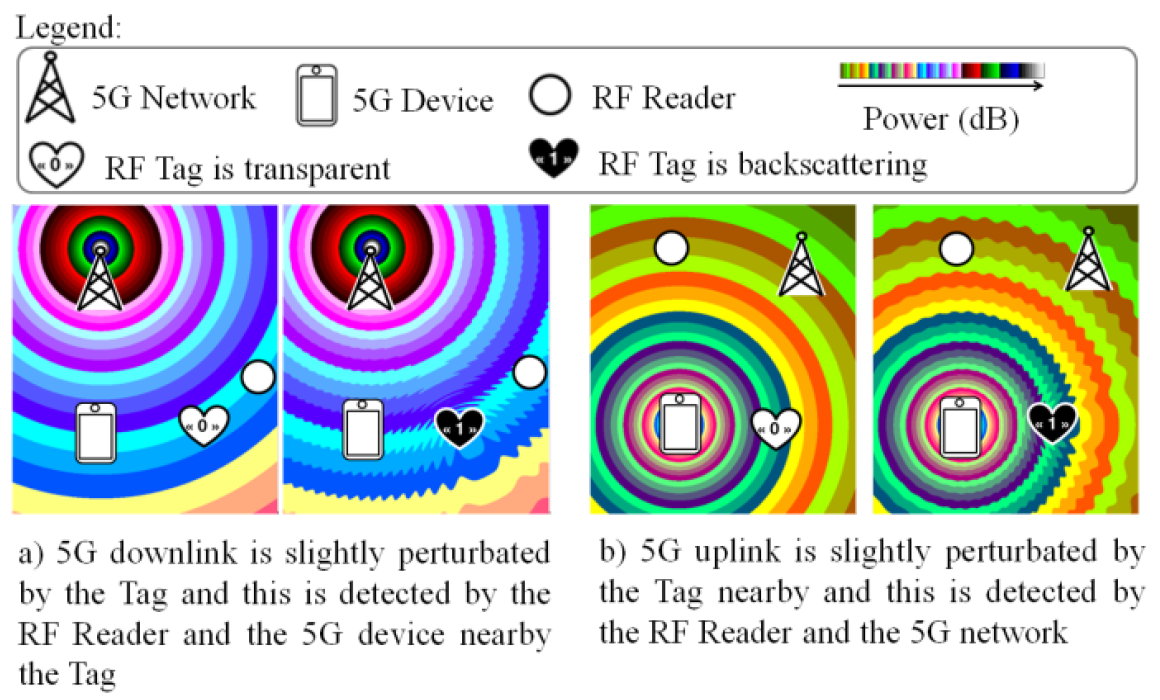}
\caption {Ambient Backscatters in 5G networks} 
\label {Ambient_Bk}
 \end{figure}

As a future operator of 5G networks, we believe that this technology could potentially help the massive development of IoT in a green manner. Indeed, as illustrated in Figure \ref{Ambient_Bk}, if applied to a 5G, the ambient backscatter concept can benefit from a large and dense population of sources and readers. Indeed, numerous 5G network base stations and 5G devices could play the role of sources. Also in addition to deploying RF readers, one could upgrade 5G devices and 5G networks with the reader capability.

\begin{figure} [ht]
\centering
\includegraphics[width=8.5cm]{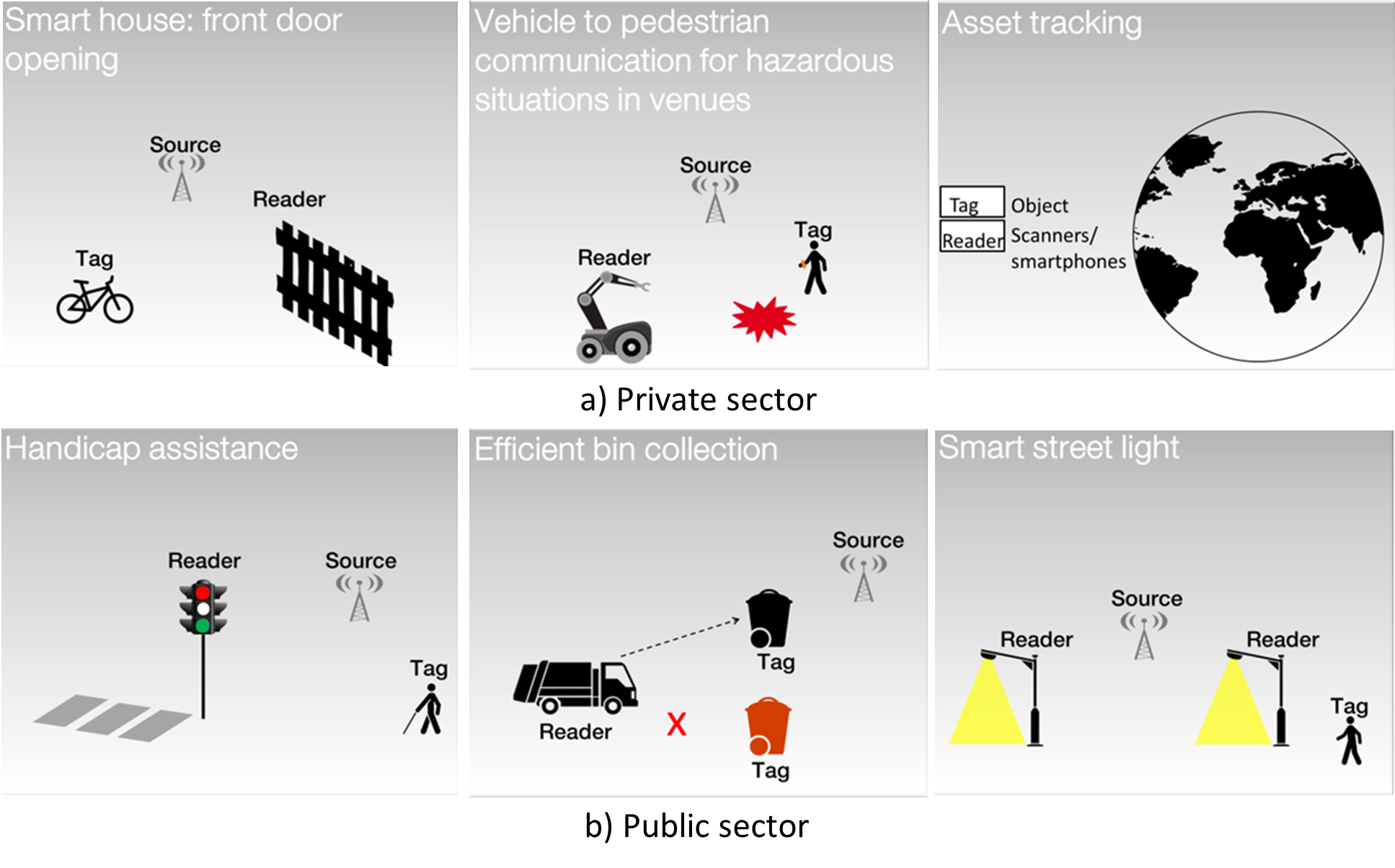}
\caption {Example of potential use cases of the ambient backscattter, with 5G network as source, and RF readers deployed on smart cities, factories and homes.} 
\label {Example}
 \end{figure}

Figure \ref{Example} illustrates potential use cases of the ambient backscatter system. In these examples, the 5G network serves as a source of RF waves, and RF readers are deployed on the environment (smart cities, smart factories and smart homes) and detecting nearby RF tags. The main idea is to render this environment “sensitive” and “reactive” to tagged objects or people.  Figure \ref{Example}-a) ilustrates examples for the private sector: a door opens when it detects the arrival of a bike, avoiding the owner to get down the bike, a fast moving autonomous moving machine avoids the collision with a tagged person circulating in the factory, a tagged package would be scanned regularly during its trip. Figure \ref{Example}-b) illustrates examples for the public sector: a traffic light warns a handicapped tagged person, a truck is informed whether a bin is full or empty and stops to collect the bin only when necessary (to save fuel), a streetlight intensifies its light only in a presence of a tagged person. All these services could be provided with existing IoT technologies, such as RFID \cite{refC}, low power wide area technologies \cite{refD} or cellular IoT \cite{refE}. However, all these technologies would additional RF waves to be generated. We believe there is an opportunity to try and offer these services with “energy-free communicating” tags that would re-cycle ambient RF waves to send their messages and that would recycle ambient energy (solar, motion or RF energy, depending on the use case) to power themselves.

\subsection{Advanced receivers} 
At the BS side basic scientific levers for green are well known and some of them have been introduced in this document. On the receivers’ side, their impact on network power consumption is not often considered since the power consumption constraint is already inherent to their design. Indeed, autonomy is a fundamental factor that is considered in the DNA of the device and therefore the degree of freedom to optimize this area is very limited. 

Also, receptors consumption is mainly driven by the screen more than by the radio part, i.e., according to studies conducted at Orange Labs receivers \cite{ref19}, energy consumption corresponds to 50\% screen, 20\% signal processing and 30\% radio. This means that the most impacting factor, i.e., the screen, is a manufacturer issue, where operators cannot intervene. The radio part could be still optimized but many power optimization software have already been implemented, e.g., UL power control, so there are left few opportunities on this side.

It is important to keep in mind that the device transmitted power is the dimensioning factor of wireless networks. The reception capabilities of the mobile and its sensitivity define the network deployment especially for sub-urban and rural areas. The receivers’ quality i.e., sensitivity, transmitted power, noise floor, is usually defined by standardization bodies. The required performance is the result of hard negotiations between operators and manufacturers with very tight tolerances. The question then is what can be done from the energetic point of view to enhance the mobile performance and the network global consumption? This question has two possible answers, 1) increasing the receiver sensitivity, which would allow reducing the BS transmitted power, and therefore saving energy and 2) increasing the receiver transmitted power in the uplink, which in theory would allow to have a higher Inter Site Distance (ISD) between BSs and therefore reducing the amount of needed sites. For instance, +1 dB transmitted at the receiver side is equivalent to around -10\% of BSs. In the first case the gain can be limited since currently there is not linear relation between the BSs radiated power and their consumed power. In the second case it implies a huge investment because sites' acquisition is very difficult to obtain. This is why, in the present circumstances, the performance of receivers is more favorable to capacity improvements in terms of QoS or Quality of Experince (QoE) than energy savings. 

\subsection{Modulation: compromise between spectral and energy efficiency}

The choice of digital modulation for a given transmission system is based on two main criteria, 1) the SE whose target is to occupy a minimum bandwidth to transmit a given amount of information with a given QoS, and 2) the EE, which includes the power required for transmission and with the recent green trend, the consumed power by the circuits. The impact of the circuits’ consumption is considered in some research papers and by national and international consortia and projects such as GreenTouch, Earth, Ecohome, GreenComm, etc.

Before examining how the second criterion (EE) intervenes and will continue to intervene, it is worth making a quick assessment of the choices made in terms of modulation for radio communication systems. In terms of modulation types, there are two main groups, i.e., the linear modulation systems and the non-linear modulations. Linear modulation systems can have a dimension by Amplitude-Shift Keying (ASK) and Phase-Shift Keying (PSK) or be two-dimensional Quadrature Amplitude Modulation (QAM). Those with non-linear modulation by frequency shift are Frequency-Shift Keying (FSK) and Minimum-Shift Keying (MSK). The former are more spectrally efficient while the latter are more energy efficient. It has to be considered that in addition to SE and EE, other selection criteria are involved, i.e., the out-of-band radiation, resistance to distortion introduced by the imperfections of the transmission systems and the channel, the cost implementation, the constant envelope property, etc. Therefore, it is a multi-criteria optimization problem which will have different solutions depending on the intended application. 

In cellular communication systems the choices from the 2G until the 4G have always targeted an increase of the SE, taking into account the constraints of non-linearity of the PA. 

\begin{itemize}
\item The GSM system provides good SE due to its Gaussian filter introduce in the form of Gaussian Minimum Shift Keying (GMSK). Moreover being a constant envelope modulation, the non-linear
amplifier of the radio front-end does not affect its performance. Naturally, being a modulation with two states, it could not meet the requirements of higher data rates.
\item In 3G the used Wideband Code Division Multiple Access (W-CDMA) transmits Quadrature Phase Shift Keying (QPSK) which means a double capacity for the same symbol period.
\item With the possibility, among other modulations, to transmit the 64-QAM (6 bits transmitted per symbol), LTE has allowed a significant capacity increase compared to previous generations. The throughput increase is higher than the allocated bands are wide, which justifies the choice of the OFDM multicarrier modulation. The disadvantage is that due to its principle of multiplexing several signals in parallel, OFDM produces high values of the Peak to Average Power Ratio (PAPR), which limits its EE. This problem can be handled in the case of the downlink, but it is not as simple for the uplink for which a different modulation scheme, the Single-Carrier Frequency Division Multiple Access (SC-FDMA), has been selected.
\end{itemize}

Other examples such as WiFi and WiMax show that for high-speed connectivity systems, the choice goes toward modulations with large number of states, e.g., 256-QAM for the IEEE 802.11ac, coupled with a multi-carrier transmission in wide frequency bands. In contrast, short-range, low speed connectivity systems use modulations with low number of states. Thus, with the Offset-QPSK modulation in ZigBee, as the I and Q channels are shifted by half a symbol time, a constant envelope property is obtained. Bluetooth, having the same constraints in terms of EE uses GFSK modulation which has very similar properties to the GMSK one.

This brief overview can also be interpreted from the fundamental principles established by Shannon in information theory that show that if we increase the spectral efficiency the EE is reduced and vice versa. Two choices in modulations with different characteristics illustrate this postulate. On the one hand, if the number of states M of a QAM is increased we obtain the M-QAM which increases its SE and decreases its EE. On the other hand, if the number of states M of a FSK modulation is increased, we obtain the M-FSK, increasing its EE and decreasing its SE.

This type of analysis, performed without considering the circuits’ power consumption, is still valid as far as it is applied to long and medium-range communications i.e., distance of around hundred meters or more. Contrary, for short-range systems, whose number is expected to grow with the small cells and the arrival of the 5G, the circuits’ consumption changes the relationship between SE and EE. In this case the EE not only depends on the energy used for the transmission but also on the one dissipated in the circuits. In the example above, the total energy for M-QAM will not increase monotonically with M, but there will be a value of M that will minimize the overall consumption. Determining this optimal point supposes the availability of sufficiently precise transmission and consumption models of the different circuits. In practice, analysis can be limited to certain key blocks. It is well known that the analog part represents the largest power consumption, while at the level of the digital baseband processing, the blocks to take into account are the Fourier transform at transmission and reception for multi-carrier systems and the channel decoding part at the receiver.

Finally, please notice that the optimization problem of the overall energy consumption does not only concern the modulation and can be applied to other physical layer blocks, particularly to the MIMO systems dimensioning. \\

\section{Network control}
\label{section5}
Nowadays, the IT-ization of telecommunication networks opens a new area of self-managed networks. The IT-isation that consists of replacing any dedicate element by a controlled and programmable element brings about flexibility and elasticity in the network management. This approach to put in place governance policies which are adapted for the operator needs (QoS, energy, latency etc.) as a function of its network specificities such as implantation, capacity, user profiles, etc.

\begin{figure} [ht]
\centering
\includegraphics[width=6cm]{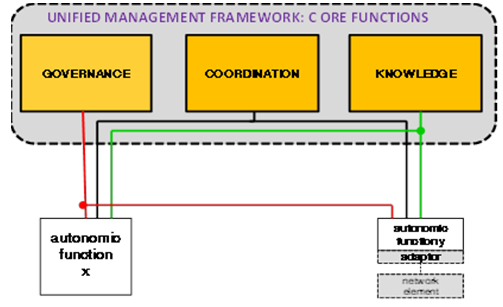}
\caption {Unified management framework} 
\label {Unimngt}
\end{figure}

Consider for example the low consumption modes. Nowadays, these functionalities operate individually and locally. A network element can turn into idle mode when traffic to be served is low or null, independently of its neighbors, the geographic situation, or the existence of other elements allowing to physically or functionally take over in case of break down/emergency recovery. This is the very basic of a green network. In fact, the introduction of Self Organized Networks (SONs), or autonomics (a system empowered by autonomic functions) allows today to have a more global view of managed systems leading to more significant gains. SON function allows acting upon a certain number of network parameters in allowing optimizing any KPI, see Figure \ref{Unimngt}.
We describe here the prerequisites for managing a “Green” network, with a special focus on the radio access network. We explain the network management functional blocks that allow taking into account these “prerequisites”. We utilize the Unified Management Framework (UMF) solution proposed in the framework of the FP7 European UniverSelf project \cite{ref20} due to its simplicity. It is noted that this solution is equivalent to the Generic Autonomic Network Architecture (GANA) standardized by ETSI AFI \cite{ref21}. The specificity of the radio access network is its complexity which is related, among others, to the different, possibly overlapping radio access technologies on the one hand, and its heterogeneity namely the coexisting macro-, pico-, femto-cell and WLAN access points on the other hand.

Among the requirements for Energy Efficient management of the network are (non-exhaustive list): 
\begin{itemize}
\item Monitoring of the energy consumption of different network segments / access technologies.
\item Update of the time profile (at different time scales) of the traffic demand per access technology, per geographical zone etc. Such profiles can be used by policies to activate rules of SON functions, to construct new rules or to refine existing ones. 
\item Introduce EE policies allowing:
\begin{itemize}
\item to direct traffic towards the most EE infrastructure (traffic steering),
\item to configure network parameters and thresholds or Radio Resource Management (RRM) parameters,
\item to activate SON functions such as sleep mode, off-loading traffic towards less energy consuming network nodes, etc. 
\end{itemize}

\item Guarantee conflict free operation of SON functions; establish activation order, priorities and time scales of SON functions and policies in general (orchestration).
\end{itemize}

The requirements are mapped onto the functional architecture of the UMF which comprises the following three Core Blocks: 
\begin{itemize}
\item Governance: managing the network using policies that control the network by means of SON functions. The policies consist of rules that translate high level objectives into low level objectives and commands for the SON functions. 
\item Coordination/orchestration: responsible for the inter-operation of autonomic functions: triggering and ordering SON functions; avoid conflicts, enforce stability and jointly optimize (possibly coupled) SON functions. 
\item Knowledge: responsible for managing information in the UMF system: collection, aggregation, storage, processing and distribution of information. Knowledge building from data for UMF needs (e.g., refinement of policies or orchestration mechanisms). 
\end{itemize}

This technology allows disposing of green governance which combines several levers. Among the SON functions that could be designed for green networking are:
\begin{itemize}
\item Turning off elements that serve low traffic and steer it to their neighboring nodes. 
\item Turn off superposing technologies (for example GSM1800 with respect to GSM 900) during off-peak hours. 
\item Modulate the activation of capacity elements as a function of traffic and maintain the connectivity.
\end{itemize}
 
\begin{figure} [ht]
\centering
\includegraphics[width=7cm]{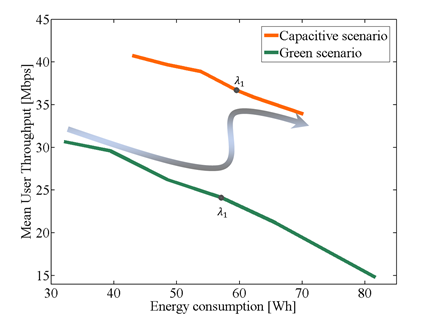}
\caption {Network with SON approaches} 
\label {SON}
\end{figure}

Let us take a practical example of how to manage the network using autonomic approaches. Starting from the daily traffic profiles we know that users’ data demand is approximately x5 during rush hours compared to night time. Access networks need then to adapt their performances to fulfill the huge capacity demand during rush hours while this capacity has to be downsized during the night. A first stage study performed at Orange Labs evaluates the use of policies that allow the network to switch from a green network to a capacity network and vice versa depending on the traffic load. For example, Figure \ref{SON} presents results for a classic macro deployment in an urban area. When traffic load achieves $\lambda_1$, results show that the best policy is to activate the capacity network. It allows increasing the spectral efficiency by a factor of 7.

This study shows that a global solution could be operated to adapt the network performances regarding the customers data demand and operator green requirements. However, such a solution would not be possible without an automatic assessment of traffic demand not only at a single node level but at regional level. Actual sleep modes, which operate at a single node level, are effective. 

Anyhow, the next step is to extend them toward global solutions that will allow a large area control without loss of QoS. A global operator policy could then specify the region, i.e., rural, urban, dense urban, and the period, i.e., night, day, rush, or even the days when activating such solutions based on traffic demand autonomics.  \\

\textbf{Network transformation} \\

\begin{figure} [ht]
\centering
\includegraphics[width=8cm]{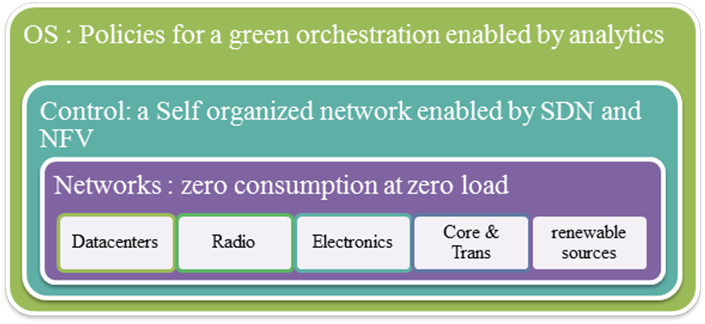}
\caption {Green network transformation} 
\label {green}
\end{figure}

For a green network transformation there have to be introduced software and hardware considerations. From the software perspective the solution should be to implement a global green controller. It would manage the network depending on the operator objectives in each scenario. Many efforts driven by Network Virtualization Function (NFV) are ongoing towards building an operating system to make networks reconfigurable. Ideally a green algorithm should be comprehensive and look for the holistic optimal performance of the network, controlling the NFV functions in order to find the tradeoff between capacity, QoS, latency, energy, etc. From the hardware perspective the objective is to have zero consumption at zero load. In what follows more details are given and Figure \ref{green} summarizes the network transformation perspectives. 
\begin{enumerate}
\item A global Operating System dedicated to green. This OS will administrate our policies and orchestrate the network operating functions in accordance with our policies. The administration functions and policies are now being developed at Orange Labs in parallel with Software Defined Networks (SDN) and NFV programs. This program should then incorporate green policies at the beginning to allow multi-objective optimization of our networks. For example, in rural areas, for wireless networks, a policy could be stated to foster EE rather than SE while, during rush hours, the contrary could be imposed during a short period.  

\item Self-organized functions are now being tested and deployed to allow an automatic control and optimization of our networks. Those SON functions have to be managed taking into account usage analytics back-propagated from customers to the network management system. 

\item Build hardware that consumes almost zero energy when no service is delivered and therefore enhance the proportionality between consumed and delivered energy. This concept was purely theoretical some years ago but is now well accepted by our suppliers especially for future networks (5G, switching and routing…). 
\end{enumerate}

\section{Adapting networks architecture towards green}
\label{section6}
\subsection{Distributed content using CDN and NGPoP}

Data centers (DCs) are technical sites with high reliability requirements. They are an important part of the networks and they represent about 20\% of the total Group technical energy consumption. Today in Orange, we have less than eighty DCs \cite{ref22}. For a long time research in this area has been focused on two levers, thermic and power distribution systems. These two levers allow acting on the Power Usage Effectiveness (PUE) which represents the consumption overhead of the site technical environment compared to consumption of ICT equipment. The works in this field, notably the ones undertaken at Orange Labs, have led to highly innovative solutions such as direct air free cooling, exothermic building etc. Notice that these solutions allow handling the problem of sites support systems consumption but they do not act in the equipment consumption level.

There are many issues regarding green DCs, among which:
\begin{itemize}
\item How to make less energy greedy ICT equipment: hardware issue related to electronics.
\item How to optimize their effectiveness: software issue related to operating load.
\item Consolidation vs distributed architecture.
\end{itemize}

Today, many levers towards green in this field are already addressed within Orange Group: metering, optimization, ICT equipment performance evolution, i.e. speed, CPU, temperature ranges, scalable hardware consumption depending on servers load. A special task is ongoing in many Orange countries to consolidate DCs, targeting to replace old ones by a reduced number of bigger DCs. This initiative is very beneficial as it will allow impacting directly the global consumption of the group through a reduced number of sites. However, Content Delivery Networks (CDNs) which are functional elements of a larger functional structure named Point of Presence (PoP) hosted in a small DC are being deployed in many Orange regions in order to enhance QoS and also to deliver dedicated services. These PoPs are now studied regarding optimal deployment. Basically, big and small DCs will coexist and an optimal solution should be studied since they pull in opposite directions. In this domain, many research initiatives are conducted essentially by INRIA, Mines Nantes. Innovative solutions such as load balancing and sleep modes are available. However, those features need to identify which services should be hosted in big DCs, which ones could be moved closer to users, or both.

Rich content and often accessed web content can be stored in CDNs. The CDN is a large distributed system of servers deployed in multiple PoPs, each in a small DC. The new generation point of presence: NGPoPs, across the Internet, will host the new virtualized CDN, vCDN. The CDN and future vCDN can help reduce subscriber churn, facilitate development of value-added services and reduce traffic on the core network. Indeed, the CDN and NGCO (see below) deployment is now acted and some traffic will be soon offloaded from big DCs. Content providers such as gaming, media companies and e-commerce vendors pay CDN operators to host and deliver their content to their audience in a more efficient way. 

At the same time, operators are in general interested in delivering the most frequently used/rich services with the less hops as possible. The idea should be then to optimize the content distribution in such a way that upper layer elements can be switched to sleep mode or at least partially deactivated as often as possible. Particularly, in \cite{ref23} the authors study two scenarios and show for both of them that important saving in terms of energy, yearly monetary and bandwidth can be achieved.

\subsection{The Next Generation Central Office}

Next-Generation Central Offices (NGCOs) or edge offices comprise the service edge of the network where service and subscriber management, routing and transport infrastructure, and customer support infrastructures reside. The Central Office (CO) is more or less the collection node of the access before transport and core networks. 

It is worth mentioning that if decentralization is on the road for the core and DCs, the access is getting more and more virtualized. Indeed, current COs are built separately for fixe and wireless networks. Those COs are now getting more and more importance as the two networks are converging and evolving. In the fixed network, the cohabitation between Passive Optical Networks (PON) and ADSL impose a new deployment optimization while in the wireless network the separation between the radio head (radio part) and BBU (processing part) open the door to new BBU hotels, cloud and virtual RAN architectures. All these options should be analyzed from an energetic point of view.

\begin{figure} [ht]
\centering
\includegraphics[width=9cm]{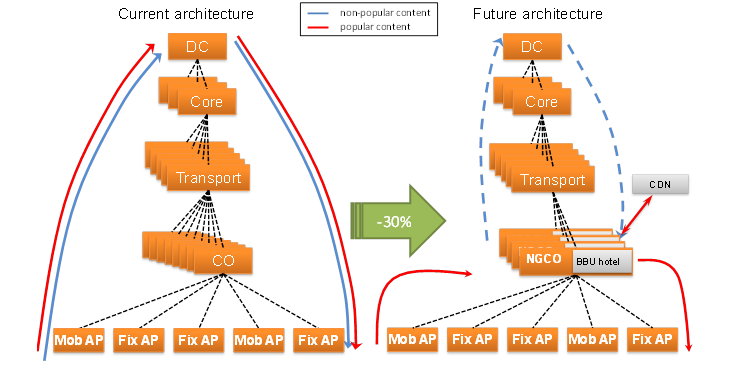}
\caption {Network future architecture} 
\label {NetA}
\end{figure}

An internal study showed that the optimum distance between broadband customers and their related CO as well as a mobile user from the BS would be about 20 Km. This would reduce the number of COs by 80\% \cite{ref24}. Figure \ref{NetA} represents the current and future network architecture. The current COs would be replaced by the NGCOs and its associated CDNs. It is worth mentioning that a CDN could be installed at national or a regional scale. As presented in Figure \ref{NetA}, some functionalities of the DCs could be descended to the NGCOs while, some functionalities currently running at the Access Point (AP) level such as ePC, v-BOX, mobile GW could be ascended to the NGCOs.

The basic idea behind NGCO and the integration of small-DCs is the possibility to serve information’s and services to our users with a very low latency in a sort of a close loop that avoid long distance bits traveling and rooting around the world. Of course, this principle would work if the most energy consuming services are the ones installed close to the user. Furthermore, if those popular services are led down in the network, part of the transport and core network would be off-loaded and put into sleep mode. 

Nevertheless, we can do a parallelism with the issue of wireless access network densification introduced by heterogeneous networks, i.e., in terms of energy consumption, is it better to have a centralized or a decentralized solution?. Namely, for the future network architecture, is it better to have a big DC (always active and far from users) or several small CDN (closer to customers and activated dynamically)?

The answer is not simple since we would probably need both architectures. Ideally, from the EE point of view, the number of active network elements that allow the user to get its information should be as low as possible. Specific studies to deal with this issue have to be launched.

In conclusion, the integration of NGCOs and CDNs in the network architecture could introduce important gains in terms of EE due to the reduction on the network components and the possibility of switching to sleep mode some network elements, arriving to a sort of flatter/decentralized network architecture. A study in Orange Labs showed a high impact on the total COs consumption i.e., a reduction of around 30\% \cite{ref25}.

\subsection{Share signaling and keep the useful traffic}

The digital traffic is the dimensioning element in terms of network capacity. Since the beginning of fixed and mobile networks it is a performance indicator and a constraint element. The traffic is always expressed with the same metric, namely the volume of bytes passing through a system, but its nature is more and more heterogeneous. For instance, we can distinguish between two types of traffic 1) user’s traffic or traffic that is requested by users and 2) signaling traffic which represents all the information exchanged for the establishment, routing and management of the communication.

The so-called “dormant” methods reduce the energy consumption of network elements when they have nothing to transmit. The idea is to turn off all or part of the network elements according to the data flow. These features are beginning to be integrated in the telecom sector but their effectiveness face important challenges i.e., even when there is no traffic there is always signaling traffic. This does not allow a full sleep mode and therefore having significant energy savings, for more details see Section \ref{ASM}.

On the other hand, the signaling is becoming more complex and traffic demanding. Much of today's signaling is produced by API-type applications, which constantly perform requests to the network for updates, keep awake and other information needed for their operation. This traffic generated by smartphones is completely independent of the user behavior (background traffic) and nowadays it consumes up to 50\% of radio resources.

Ongoing researches since 2011 study the feasibility of separating user data and the signaling. The interest of doing so is large:
\begin{itemize}
\item For green: separation allows to switch off not necessary traffic elements and maintain the signaling traffic required to maintain connectivity.

\item A new business model: separation allows sharing signaling between several operators, and even the emergence of new actors such as coverage operators working in B2B mode. Whereas the data network is managed by traditional operators. This idea would allow sharing the network while maintaining the competitiveness since traditional operators keep control of their traffic assets and therefore manage the network capacity.
\end{itemize}

Almost all manufacturers agree on this idea. In terms of green it means to have BSs without signaling. In theory, this would allow to completely switch off these BSs in the case where there is no user traffic demand. In other words, the station consumes zero Watts when traffic is zero. This idea emerged in the GreenTouch consortium. These BSs are called ``slave" and operate in synchronization with BSs called ``master" which provide users connectivity. The first models will come out in less than two years for small cells since their use is consistent with this principle and furthermore, they reduce the interference. Finally, this idea will take little time to pass from fundamental research to the field and the arrival of “phantom” small cells will allow judging the operability of such a solution. The studies performed within GreenTouch have shown that such a solution coupled with the sleep mode can reduce the overall consumption of the RAN by a factor of 6 compared to a conventional network with 2020 traffic model. This technology could be easily implemented for macrocells but it requires further studies for its deployment since studies presented by GreenTouch showed that it could not be efficient \cite{ref26}. \\
%
%
%

\section{Data centers, servers and software}
\label{section7}
In the mid-90s the internet became of public use. The first Asymmetric Digital Subscriber Line (ADSL) deployment in France dates from 1997. Among many other things, the success of this technology has been given by the combination of two factors, i.e., the very low oil prices ($ 10 to $ 20 per barrel) associated with the Moore's law, which states that the computational capacity of microprocessors doubles every 18 to 24 months. Therefore, up to now, all the internet assets (e.g. PCs, terminals, ICT equipment, software developments, DCs, etc) have been designed and developed without strong constraints in terms of energy consumption.

DCs encompass both ICT equipment it hosts and technical environment required for hosting them. Due to the reasons mentioned above, their consumption is impacted at several layers, each one magnifying the inefficiency of the lower one.

\subsection{Software eco-design}

The chipset computational and information processing capacities have evolved in such a way that they seem “endless”. This has led to the development of a very rich programming technology but also highly resource consuming. Namely, Intel continues to develop more powerful microprocessors which allow Microsoft building up richer software releases. For instance, Windows Office 2000 can run on a Pentium 75 MHz whereas Windows Office 2007 requires at least a 500 MHz processor. For instance, the package Windows 7 + Office 2010 professional requires 71 times more memory, 47 times more disk space and processors delivering 15 times more processing capacity than the Windows 98 + Office 97 package, to deliver similar services such as text editor, slides editor, spreadsheet, overloaded with frivolities and functionalities scarcely used. The requirement for new processor functionalities and therefore faster processors by latest software suites leads to an accelerated obsolescence of the “old generation” electronic devices and a significant amount of Waste Electrical and Electronic Equipment (WEEE). From the software development side, the situation is the same and causes a drastic growth of energy consumption by networks software elements and especially by DCs.

Furthermore, new coding languages (such as Java, Node.js, etc) offer rich features but have very poor capabilities regarding processors resource management compared to low-level languages such as assembler or C. With these new programming languages, if there are resources constraints the solution is to find more powerful equipment. That is why it is very common to come to situations were old computers have difficulties accessing heavy content websites.

Additionally, the co-evolution of the smart phone Operational Systems and the CPU computational capacity should guarantee ability to support even poorly designed applications capable of drying the batteries quickly. The introduction of more powerful Gateways and Set-top Box processors as well as stacking software that associated with use cases like Any Time Any Where Any Device (ATAWAD) constrain the use of low consumption modes, leading to have devices active 24/7 that are used only few hours per day.

All these examples confirm what Niklaus Wirth had mentioned in 1995, “Software is getting more rapidly slower than hardware becomes faster”. We can then talk about the “Bloatware”. All those heavy software’s can be seen as a technical debt that we try to compensate throughout its life cycle.

Also a significant reduction of the ICT environmental footprint will depend on the development of simplified hardware, implying an important reduction in the energy consumption, which goes against the current momentum. Initially, this could be feasible only by an awareness-raising of the developers on the constraints of energy consumption of the applications/software they develop. To do this, tools to monitor the application/software energy consumption should be provided. Then, the second step is to provide developers with recommendations on more sober patterns. This requires identifying the right coding style for different technologies following “green patterns”.

Indeed, software eco-design can be applied to all coding technologies, so it is irrelevant to catalog technologies in a greener scale since each one has its own use and field of application. Also, it is important to drive studies for each coding technology to identify the impact of the main sequences implemented by developers in terms of power consumption.

Finally, an important axis in eco-software design concerns the functional outline of the application to be developed. An application containing multiple functionalities will be more complex and therefore the consumption of the underlying resources will be more important. To have sober applications, it is hence essential from the marketing side to consider the relevant features for the customer instead of introducing “just in case” features.

Studies in Orange Labs showed that by simply replacing high energy consuming software sequences by sober ones energy consumption can be reduced by 7\%. Furthermore, if improvements were introduced at the architectural level using optimized libraries in addition to the optimization of applications’ functionalities among others, energy consumption could be decreased by up to 40\%, as achieved for Orange application Business Everywhere.

As a conclusion, for an actor such as Orange, software eco-design can introduce important savings in terms of OpEx and CapEx, i.e., applications consuming fewer resources (OpEx) and increasing lifetime of its equipment (CapEx). From the users’ perspective, benefits are the same, adding the advantage of more user-friendly applications.

\subsection{Ongoing studies in cooling}

Energy consumption by cooling systems for hosting ICT equipment is a major issue for operators and over the top companies. A cooling system can consume between 20\% and 50\% of the total power of a site.

Standards have helped establishing norms to set the temperature and humidity operating ranges. Telecom equipment must normally work meeting these norms in order to achieve these ranges. Recent studies \cite{ref28, ref29, ref30} show that it is possible to extend the operating ranges of climatic equipment while maintaining normal operation.

\begin{figure} [ht]
\centering
\includegraphics[width=9cm]{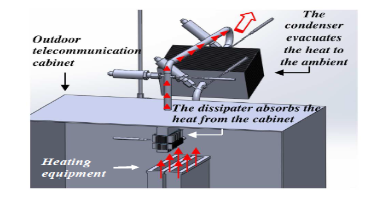}
\caption {Caen university prototype (project Opera-Net2)} 
\label {caen}
\end{figure}

Studies \cite{ref31} on thermal storage at the building structure show that it is possible to compensate the variations of the external temperature. This can be achieved by the use of heavy materials, phase-change products and an optimized air flow. Several patents \cite{ref32, ref33} on this subject have been presented, one of them is showed in Figure \ref{caen}.

These limits are used to define the target temperatures for air conditioners or heaters. One of the first things implemented was to extend the operating ranges to profit of bigger operational ranges.
Conventional air cooling systems are highly energy greedy and getting close to cooling capacity limits. One possible way of achieving substantial improvements is the use of another coolant, e.g., water, oil, diphasic liquid, etc.

\begin{figure} [ht]
\centering
\includegraphics[width=5cm]{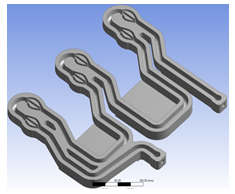}
\caption {Water circuit presented by Nokia in the framework of operanet2 project} 
\label {water}
\end{figure}

Liquid cooling technique involves setting up a heat exchange interface between the ICT system dissipating heat and the liquid that will absorb this heat. This technology could be even integrated directly to the electronic card following strict design constraints so that the liquid flows as closer as possible to the card components, see Figure \ref{water}. It is needed to conduct research on coolants with high thermal capacity and low viscosity. The fluids should be studied to identify their physical properties regarding their performance. They must be dielectric to prevent short-circuits at cards, non-corrosive and environment friendly. Two-phase fluids should be studied in order to increase the heat transfer performance.

\begin{figure} [ht]
\centering
\includegraphics[width=6cm]{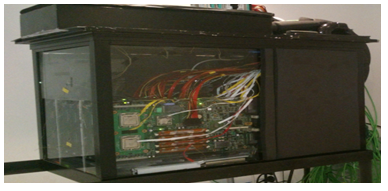}
\caption {Immersion cooling container prototype in Orange Labs} 
\label {cooling}
\end{figure}

Immersion technique consists in immersing servers in oil or other dielectric coolant, as presented in Figure \ref{cooling}. The oil has the advantage of not being volatile but it is viscous, therefore, the pumps must be properly dimensioned. The use of a phase change product, such as 3M, must be mastered in the fluid evaporation. 
  
\subsection{Energy storage}

France was pioneer with the battery invention of Gaston Planté. It has more than a dozen research laboratories working in this field from which 80\% perform basic research. Notably, we find the University of Amiens with world class advances thanks to the work of Professor JP Tarascon (College of France) and Professor Armand inventor of lithium metal and iron battery LiFePO4. 

Orange interests in this domain are not related to fundamental research in electrochemistry or in the development of mobile devices battery, but in the applied R$\&$D to reach mass production of low cost advanced batteries with alternative technology to be introduced for network energy backup. 

In some countries, due to energy transition policies that are being implemented, it is expected an important increase in their storage performance and capacity. With the use of smart grids, during peak periods interesting economical interactions can be achieved i.e., peak consumption adjustment using renewable energy and batteries.

It is very important to closely follow this subject since the active research on storage can lead to very significant breaks. To mention some examples, start-ups IonWatt, spin off of the CNRS laboratory in Rennes, working in the Redox Battery, an organometallic solution also CEA’s start-up.

Finally, it is worth mentioning TESLA marketing breakthrough with its launched Lithium-Ion battery (April 2015), the Powerwall, announced as the solution for energy storage for a sustainable home at very low prices. This battery will be produced by the future Tesla-Panasonic Gigafactory. It will charge using electricity generated from solar panels, or when utility rates are low, and power the home in the evening. Hard competition is engaged with other competing technologies e.g. advanced Lead-Carbon, NiZn and Lithium-Iron.

Orange recommendations in this area are: 

\begin{itemize}
\item The study and introduction of distributed and decentralized energy generation. It should bring important security and economic advantages since 1) it guaranties operators to give a reliable and resilience service, independently from local energy grids and 2) it avoids the energy losses in transmission and distribution, which range from 8\% to 15\% from the power plant to consumers. 

\item The use of microgrids which would enable the integration of different renewable sources of energy, energy storage, and demand response. The growing interest on the green city concept and the recent launch onto the market of more affordable batteries and solar panels from companies such as Tesla and SunPower show that it is the good timing for the study of this solution.

\item Introduce a demand response program at Orange group. It could derive considerable benefits by reducing the effective consumption by turning off some equipment, consuming its stored energy or using its renewable energy sources during peaks. Studies carried by Orange Labs showed that with the introduction of such a program in the French network the group could gain several hundred k euros /year.
\end{itemize}

%
%
%
%

\section{Mathematics for green}
\label{section8}
During the last years there has been a reconcile between mathematics and information networks which is probably due to the increase on the degree of freedom offered by the new control systems in the telecommunication networks. This trend will continue even with the Network Virtualization Function (NFV), the separation between the control and the data planes and the IT-ization of the network elements. These features would make the network more flexible and most important, programmable. This is called networks orchestration.

Mathematic allows to develop models and to optimize the network performance from input data which are the orchestration parameters. Metadata or big data sequentially or continuously feed the mathematical models. 
\subsection{Game theory for assessing services consumption}

Game theory studies strategic decision making processes. Based on mathematical models, it studies conflicts and cooperations between intelligent rational decision-makers. Initially, it was introduced in economics to understand the economic behaviors of firms, markets, and consumers. Today, however, game theory applies to a much wider range of disciplines such as political science, psychology, computer science, and biology. The games studied in game theory are well-defined mathematical objects. The following elements have to be defined: the players of the game, the information and actions available to each of them at each decision point, and the payoffs for each outcome.

Game theory is a mathematical field that has attracted for several years deep interest from communication network researchers. Notably, two application examples stand out, i.e., the medium access games for 802.11 WLAN and the power control games in CDMA systems. Game theory continues being considered to solve decision making problems in the network and there is an active research to apply it on routing, security, interference control, resource allocation, etc.

\begin{figure} [ht]
\centering
\includegraphics[width=9cm]{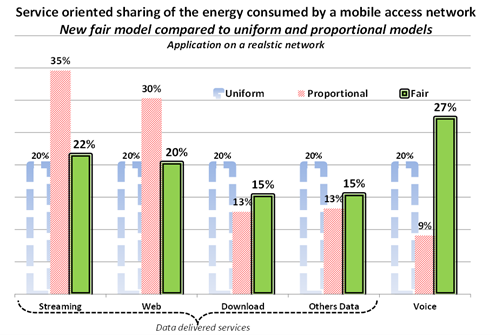}
\caption {Service consumption assessment using game theory} 
\label {sharing}
\end{figure}

Research Studies in Orange Labs used game theory, particularly the collaborative games, in order to define a fair sharing of the energy consumption of the different services delivered by its network. The main aim of these studies was to find levers to lower the global energy consumption based on hardware characteristics, features and configurations. Recently, the focus has been extended to the assessment of the impact of the different services on the energy consumption in order to reduce it. First results are already available, as presented in Figure \ref{sharing}.

The energy consumption of a given network can be split into a load-dependent part that is mainly related to traffic transmission, and a fixed part given by the consumption of the equipment in a standby mode (no data traffic). The fixed part may represent a huge chunk of the energy consumption, e.g., the fixed consumption of the mobile networks access part represents about 50\% to 80\% of its total consumption. While the energy repartition of the load-dependent part between services can be done based on the traffic consumption, the repartition of the fixed part is still an open question. A solution could be sharing uniformly the fixed part between the services and another solution could be to use the traffic proportion of each service as a basis for the energy sharing. Both methods could be considered as unfair depending on the service, e.g., services like games that generate a tiny fraction of traffic would prefer a sharing based on the proportion of traffic contrary to big players like streaming services who would rather prefer a uniform sharing as the fixed part that does not depends on usage.

To solve this problem, Orange Labs carried out some collaborative research on a sharing method based on coalition game concept, the Shapley value. In game theory, the Shapley value, named in honor of Lloyd Shapley, who introduced it in 1953, is a solution based on cooperative game theory. Five service categories are considered according to their traffic volume, i.e., three large players: streaming, web browsing and file download, and two smaller ones: voice and other minor services. Figure \ref{sharing} compares the energy consumed by the different services when using Shapley-value-based, named Fair in the figure, the uniform sharing between the different service categories independently of their traffic volumes and a proportional share which follows the volume proportions.

This result show that Shapely-value-based method is more robust towards small services since it reduces their shares in comparison to the uniform approach and fairer towards larger services in the sense that it reduces their shares in comparison with the proportional case.
This example shows how mathematical tools could be useful in the green research field to assess services energy consumption. Particularly, Orange Labs is collaborating with different actors in this field in France, i.e., Insitut Telecom and Mathematical and Algorithmic Sciences Lab of Huawei Paris. 

\subsection{Traffic analytics for energy management}

Traffic analytics for QoE management are a set of indicators measured from real traffic flows. Those indicators are currently used by operators to enhance the QoE and to detect the optimization points that would improve the global QoS. Analytics can then be viewed as a feedback flow from users to the network that allows the operator to adjust and optimize the network parameters and controls.

Orange Labs started a dedicated study in collaboration with Orange-Cameroun in 2014. This study has focused on network energy failures and the induced traffic losses in wireless access nodes. Based on Call Digital Records (CDR) harvested during one year communication, the study showed that it is possible to secure APs in a form of clusters rather than individually. It turns out that:
\begin{itemize}
\item Energy shortage events are not related to the level of traffic conveyed by the BSs (long term incidents can occur on “large” or “small” BSs). 
\item Large cities, such as Douala or Yaoundé, experienced about 10 days (cumulated on one year) of unavailability due to energy shortage reasons.
\end{itemize}

\begin{figure} [ht]
\centering
\includegraphics[width=9cm]{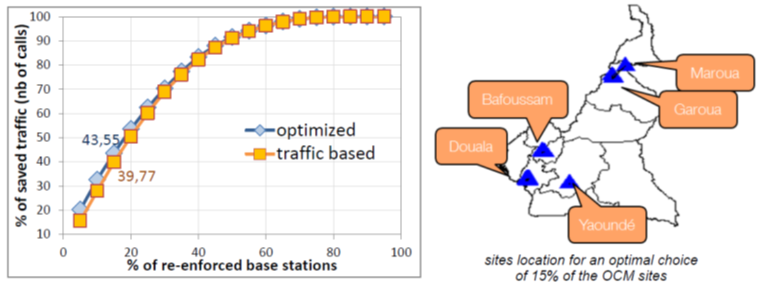}
\caption {Re-enforced BS energy supply} 
\label {supply}
\end{figure}

Also, the study showed that re-enforcing the BSs energy supply in a 15\% allows saving about 40\% of the traffic that would otherwise be lost. Figure \ref{supply} shows that this value goes up to 43\% if the BSs are chosen optimally thanks to an appropriate optimization model.

Many research studies are ongoing trying to take advantage of those analytics to optimize ICT networks in terms of energy consumption. The main advantage of analytics is that it allows a global optimization, rather than a local one. For example, balancing the EE with the SE during low traffic periods is one of the main applications of this approach. Of course, optimization could be done locally but network elements are linked and a global optimization would allow avoiding conflicting interactions. This new research field should be addressed in collaboration with big-data experts, network control engineers and SON suppliers. \\

\subsection{Reinforcement learning} 
Reinforcement learning (RL) consists on learning how to map situations to actions in order to maximize a numerical reward \cite{RL}. Several methods enable to solve a RL problem such as Q-learning which is a model-free approach where the \textit{agent} behaves randomly without any specific policy.   \\
We applied this approach in \cite{qlearning} in order to orchestrate the ASMs according to the requirements of the network operator in terms of energy consumption and delay. The control agent has to decide how many times to repeat each sleep mode level depending on the defined utility. This utility corresponds to the reward that the agent gets after choosing an action. It can be defined as follows: 
\begin{equation}
		R = - \epsilon D + (1 -\epsilon) E_{gain}
	\end{equation}
where $D$ is the delay added due to the sleep policy, $E_{gain}$ is the energy consumption reduction when using the sleep modes and $\epsilon$ is a normalized weight ($\epsilon \in [0,1]$) that denotes the importance given to the two factors $D$ and $E_{gain}$. A small $\epsilon$ means that the EC reduction is prioritized over the delay and vice versa.  

\begin{figure} [ht]
\centering
\includegraphics[width=8.5cm]{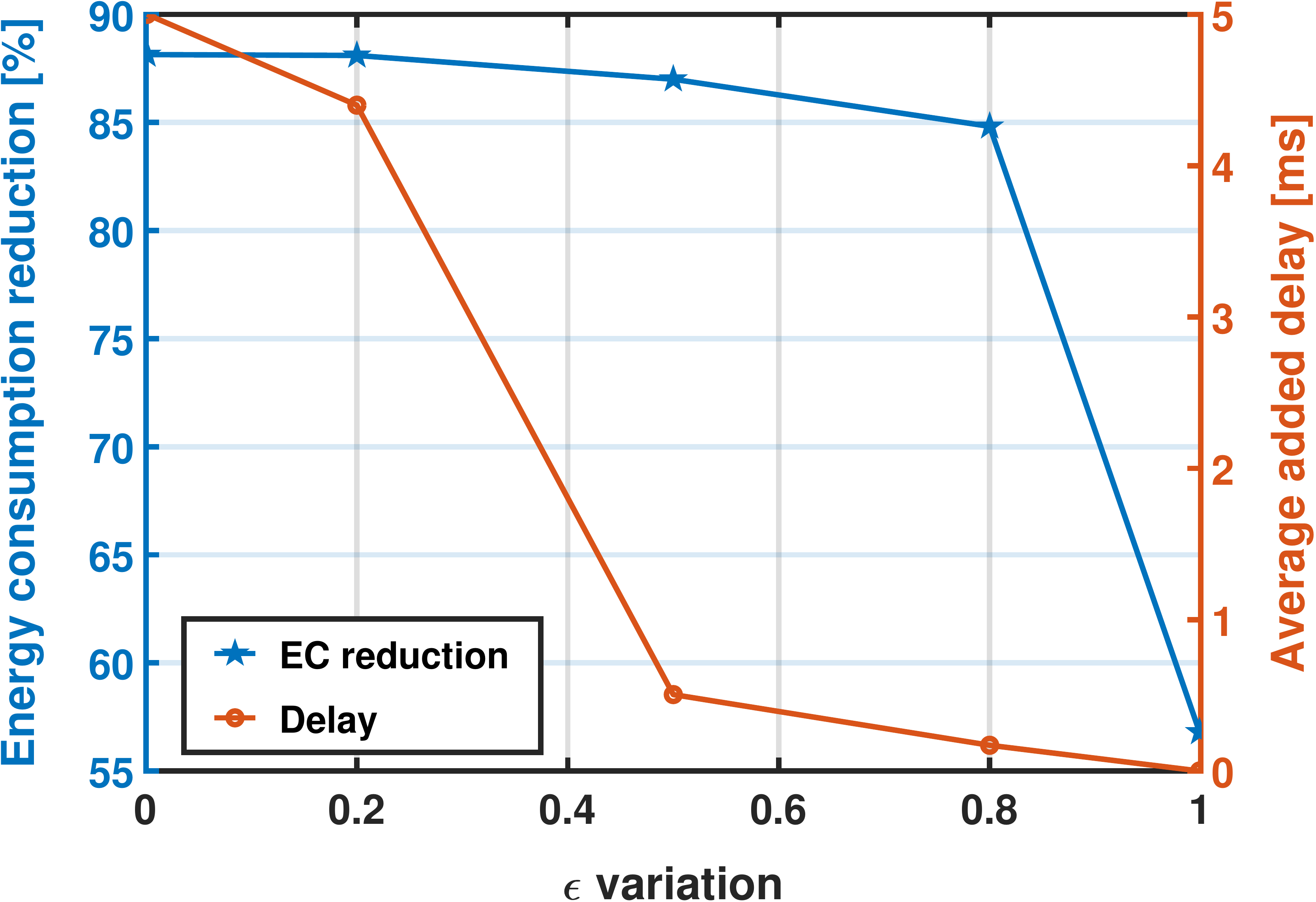}
\caption {Performance assessment of the selected policies during the exploitation phase} 
\label {fig:policies}
 \end{figure}

Figure \ref{fig:policies} shows that if we have a high constaint on the delay (high $\epsilon$), we can reduce the energy consumption by 57\% without having any impact on the delay. Whereas, when there is no constraint on the delay (low $\epsilon$), we use more the sleep modes, especially the deepest one, and the energy saving is up to almost 90\%. So, depending on the requirements of the different 5G use cases, $\epsilon$ has to be chosen carefully.  

\subsection{Stochastic process}

A Markov chain is a stochastic process describing a sequence of random changes of state in a given space of states. Markov chains are memory less processes, i.e., the next state depends only on the current state and not of the sequence of events that preceded it. Markov chains are heavily used to model a plethora of phenomena in the real world.

Queueing theory is the mathematical study of queues. This research field was introduced in 1917 thanks to the work of Erlang, a Danish engineer specialized on telephone networks management. Queuing theory is mainly used in performance analysis and systems optimization.

\begin{figure} [ht]
\centering
\includegraphics[width=9cm]{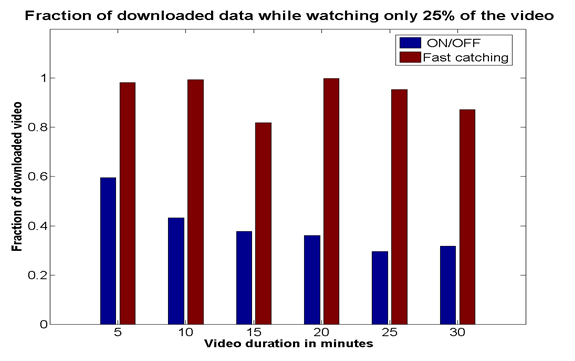}
\caption {Fraction of downloaded data while watching only 25\% of the video} 
\label {dwld}
\end{figure}

To optimize battery consumption, smartphones manufacturers use different downloads management algorithms for video services. Nowadays, Android-based smartphones use an ``ON/OFF" mechanism where the video player switches between ``ON" state when the smartphone simultaneously downloads and plays the video and ``OFF" state when the smartphone stops the radio interface during predefined time and reads the data stocked in the buffer during the ``ON" state. This strategy is designed to take into account high abandoning rate of video viewing, i.e only about 35\% of video are downloaded entirely by the viewers. Therefore, given an adequate parameters of the ``ON/OFF” periods, this strategy can reduce the transmission of the unnecessary bytes, hence, reducing the battery consumption at the smartphone side and the power transmission at the network side. Figure \ref{dwld} shows a comparison of the average percentage of downloaded bytes between the ON/OFF strategy and fast caching strategy (Normal downloading without OFF period) while viewers for different video lengths abandoning the video after watching only 25\%. In this simulation, we set the ON period to 15 seconds and the OFF period to 50 seconds.

However, a bad parametrization of the “ON/OFF” periods might lead to several video freezes due to empty buffers during the ``OFF" periods. 

Orange Labs teams are working on the mathematical modelling and optimization of downloading strategies, in particular the ON/OFF mechanism. Markov chains and Queuing theory are used, respectively to model the users' dynamics inside the cell and the dynamics of the buffer size of the video player.  The combination of both mathematical modelling allows us to optimize the durations the ON/OFF periods i.e., energy consumption and the impact on the QoE. 
\subsection{Dynamic programming}

Dynamic programming is an algorithmic method to solve complex problems by breaking them down into simple sub-problems. It relies on a simple principle, any optimal solution leverages itself on sub-problems solved locally in an optimal way.

Currently, the terminals connect to the access network that offers them the best connectivity. This policy could induce significant overload on a particular access network. This raises the question of how to distribute traffic between different customer access networks in order to reduce the energy consumption guarantying an acceptable QoE. Studies to focus on building models based on dynamic programming in order to solve this issue by allowing the design of decision algorithms that can dynamically and optimally, from an energy and QoE point of view, allocate real-time traffic flows to different access networks.

Dynamic programming is expected to have a significant impact on the development of an Operating System (OS) to the green. This area of research provides an opportunity for network operators to optimize in a comprehensive manner their networks without moving the energy consumption from one sector to another. This question is increasingly present due to the conflicting opinions regarding the effects that the cloud or the network virtualization could have. Dynamic programming may be the only means to guarantee operators to optimize their network from end-to-end.

If today the initiatives in this area are rather in a research phase, the introduction of Networks Functions Virtualization (NFV) and Software Defined Networks (SDN) in all network elements in order to make them programmable have speed up the considerations in this sense. A comprehensive operational system (Global OS) with a green feature should be conceived since now in order to place this option in the genes of future network OSs.

\section{Conclusion}
\label{section9}
It is fair to say that the evolution of ICT networks is facing a huge exponential traffic increase which naturally induces a constant growth of its energy consumption. Furthermore, deploying new networks without decommissioning legacy ones, plus the increase of the number of customers asking for always more data rates will definitely make the green challenge more complicated and challenging. We have then to admit that ICT energy consumption will probably grow in the next 10 years if an aggressive green transformation it is not launched. Research works should limit this constant growth and give us the basis of future greener networks conception, operation and maintenance. For this purpose, we should prepare ourselves to: 
\begin{itemize}
\item ­Impose green to our suppliers as a mandatory target. This would not be possible without a large consensus among the main worldwide operators.
\item ­Co-innovate with our industrial eco-system to fulfill our green requirements.
\item ­Buy specific green developments from our suppliers as they are ready.
\end{itemize}

Through this paper we have highlighted some technological levers that would allow accelerating the transformation towards green networks. In a summary: 

\begin{itemize}
\item \textbf{Fundamentals} show that combining the Shannon limit of telecommunication and the Moore’s law about electronics computational improvement may conduct us to move from a concept that we would name one-big-to-one-big towards many-small-to-many-small. In other words, green and sustainable would be done by moving from point-to-point concept to massively parallel communications that is already observed with for example Titan-MIMO in wireless communication and highly parallel computing in super servers. 
\item \textbf{Semiconductors} analysis indicates that Gallium is going to swap silicon in most of telecom applications and especially for radio equipment. This material is well suited due to its great performances for high voltage and high frequencies, it also will facilitate the integration between photonics (AsGa) and electronics (GaN) on the same chip. The cost of Gallium versus silicon has significantly decreased (from x4 to x1.5 during the last 3 years). This material could be now integrated in core and transport networks. Extension to the access network which is two orders of magnitude bigger than the core in terms of network elements will be appropriate as soon it will be economically viable.
\item \textbf{Optical technologies} have already permitted to significantly decrease energy consumption of ICT networks (FTTH for Access and optical routing and switching for Core). Optics, thanks to new components and modulations still has a huge potential progress in terms of EE. In order to go further at the access networks some initiatives should be fostered: 
\begin{itemize}
\item	Making optical amplifiers supporting rapid switch-on/switch-off. 
\item	Inventing “photonic energy consumption” scalability 
\end{itemize}
\item \textbf{Radio networks} must be particularly focused because of its tremendous energy consumption induced by multi-layer technologies that are installed and maintained even if some of them transport almost zero traffic. The first age of green wireless networks is already improving its efficiency as following :
\begin{itemize}
\item Increase the hardware efficiency by moving to GaN component technology and introducing advanced software sleep modes that should allow achieving almost zero consumption at zero traffic. 
\item Shift to massively parallel antennas with focusing capabilities. High gain antennas will allow reducing drastically the radiated power without a loss of coverage requirments.
\item Design advanced system with energy-free communicationg tags.
\item Build a global green Operating System (OS) which is necessary to balance between energy and spectral efficiency depending on the zone (rural, urban), the context (social events) and the time period (day, night).
\end{itemize}
The second age of green wireless networks would probably be more disruptive. Think that many degrees of freedom (e.g. Signal processing, communication protocols, antenna systems ... etc) have been revisited and optimized while the cellular concept is still alive since 1947. The next generation of green wireless communication will for sure change the cellular concept to user-centric wireless network. These potential breakthroughs pave the way to the principle of “always available on demand” and breaking the cellular concept.
\item \textbf{Network control} will allow operators to globally control in real time their networks. Up to now, this super-controller is being built to manage the traffic flow depending on the targeted quality of service. However, as green is introduced as an important KPI, an ad-hoc orchestrator is necessary to manage conflicting KPIs by setting an optimum configuration. For example, rural areas which represent 70\% of the wireless network consumption should be driven by EE while dense urban areas by spectral efficiency. Those studies are now being launched and should be accelerated in the next 3 years.

\item \textbf{Network architecture} should be simplified and the migration from numerous legacy central offices to a limited number of NGCO/NGPoP hosting CDNs must be seen as an opportunity to be boosted. Additionally, the right positioning of the contents inside our networks which is still an open question should be studied via research studies leveraging our data mining capabilities and mathematic tools in order to well-balanced energy consumption depending on the content location and popularity. The challenge is to find the right equilibrium between energy gains, latency requirement and business revenue.

Nevertheless, an end-to-end separation between signaling and data planes could give operators a new opportunity by sharing their signaling networks to decrease their energy consumption and maintain networks always alive.  
\item \textbf{Data centers and software eco-design} The fast software evolution has led to higher hardware resources consumption (e.g. last 10 past years memory requirements for Windows 9 vs Win3 has increased by 100). Future software versions and applications should be designed from a greener perspective. To achieve additional energy savings, the use of dynamic VM consolidation algorithms shall maximize the number of inactive servers and by the way putting into sleep mode unused resources.  Moreover, oversizing should be restricted to needs for disaster recovery plans and operations. Research should be launched soon to help operational for managing this important issue. 

\item \textbf{Mathematics for green} seems to be one of the most promising domains that should be considered in the near future research trends. Mathematics are suitable for different domains like:
\begin{itemize}
\item Modeling end-to-end energy consumption taking into account input data uncertainties and controlling their response probability. Stochastic meta-models allow overcoming this issue by controlling the response accuracy depending on the quality of the learning inputs.
\item Managing conflicting KPIs through a green orchestration brain is the fundamental building bloc of future networks. Orchestration will need multi-objective optimization tools for decision making allowing a non-aggressive and smooth network control. 
\item Pay to play: Future trends are towards assessing end-to-end services energy consumption sharing induced by over-the-top services (google, facebook, tweeter, ..) that may account for 80\% of ITN induced energy consumption in 2020. Game theory will allow a fair assessment of services energy consumption between different players. Those studies could even been extended to give our users their ICT environmental footprint induced by their usage.
\end{itemize}
It is fair to say that the given technological levers will have different positive impacts depending on their network segment applicability. For example, the ones applicable to the access networks which represent globally 70\% of the ITN consumption will be more energy efficient than those applicable only to the core or transport networks. Hence, putting all the pieces together, we have weighted and mapped the proposed technological levers impact on the global ITN consumption picture in Figure \ref{fig:summary}. 
\begin{figure} [ht]
\centering
\includegraphics[width=9cm]{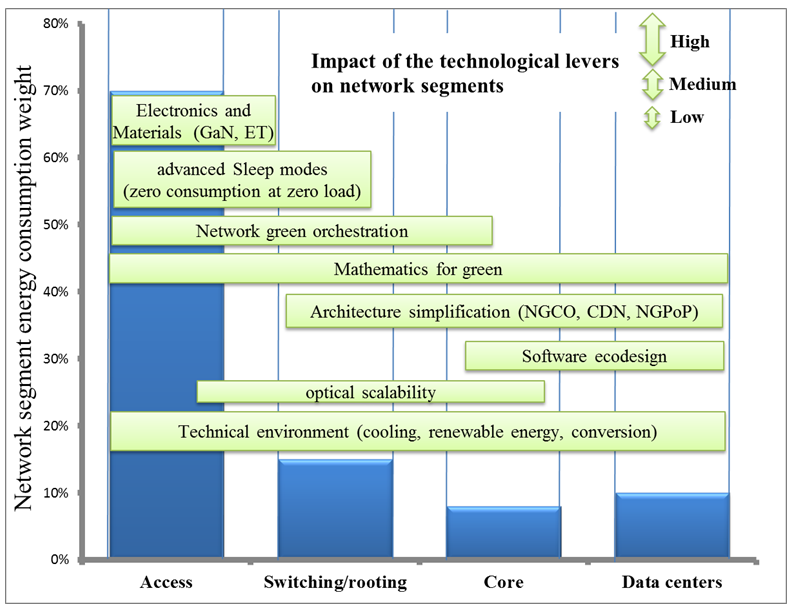}
\caption {Technological levers energy impact on network segments} 
\label {fig:summary}
\end{figure}
Finally, the green domain appears at the intersection of multiple scientific disciplines (mathematic, physic, informatics and economy). It is becoming part of all ICT activities which could be a differentiator criterion for Orange against its competitors due to its research capabilities.  Currently, sustainable development is being thought at schools as an independent course and it will definitely become an engineering ICT branch. Some universities have integrated in their masters’ program a green ICT option and operators should support this initiative through our research and development activities. Therefore, betting for a deep and real green strategy appears as the right path according to ongoing reality.
\end{itemize}

\section*{Acknowledgement}
The authors would like to thank Dr. Nissem Selmene for her work on the design of the Ambient Backscatter at Orange Labs during April 2017 to March 2018.


\begin{thebibliography}{11}

\bibitem{ref1}
	ITU (2009) ICTs and Climate Change, background paper for the ITU Symposium on ICTs and Climate Change, Quito, Ecuador, 8-10 July.

\bibitem{ref2}
	B. Nordman, S. Lanzisera, ``Electronics and network energy use: Status and prospects", 2011 IEEE Intern. Conf. Consumer Electron. , pp 245-246.
	
\bibitem{ref3}
	D. Kilper et. al., ``Power Trends in Communication Networks" in proc. IEEE Journal of Selected Topics in Quantum Electronics, Vol. 17, No. 2, 2011.

\bibitem{GreenT}      
https://s3-us-west-2.amazonaws.com/belllabs-microsite-greentouch/index.html

\bibitem{Earth}      
M. Gruberet al., ``EARTH - Energy Aware Radio and Network Technologies", in 2009 IEEE 20th International Symposium on Personal, Indoor and Mobile Radio Communications, Sept 2009, pp. 1–5.

\bibitem{SoGreen}      
https://soogreen.cms.orange-labs.fr/soogreen

\bibitem{OpNet}      
Celtic-OperaNet, www.celticplus.eu/project-opera-net2

\bibitem{ref4}
	M. Babacan, ``Lobbying and Growth: Explaining Differences among OECD Countries", Topics in Middle Eastern and North African Economies , Vol. 12, No. 1 , September 2010.	
	
\bibitem{ref5}	
	O. Renais et al., ``Migrating to a Next Gen WDM core network", the 13th International  Telecommunications Network Strategy and Planning Symposium, 2008. 
	
\bibitem{ref6}
Vlasov, Y. ``Silicon CMOS-integrated nano-photonics for computer and data communications beyond 100G", Communications Magazine, IEEE, volume 50, issue 2, pages 67-72, 2012.

\bibitem{ref7}	
	B. J. Offrein, ``Silicon Photonics for the Data Center," in Optical Fiber Communication Conference, OSA Technical Digest (Optical Society of America, 2015).
	
\bibitem{ref8}
	E. Bonetto, et al., ``Facing the traffic explosion in metro transport networks with energy-sustainable  architectures", Photonic Network Communications August 2015, Volume 30, Issue 1, pp 29-42.
	
\bibitem{ref9}
http://www.ieee802.org/3/az/index.html
	
\bibitem{ref10}	
	Orange requirements book, internal report.
\bibitem{ref11}
  https://www.celticplus.eu/project-operanet/

\bibitem{ref12}  
  ``Power Modeling of Base Stations", 5GrEEn Summerschool Aug. 2014

\bibitem{ref13}    
	Green Touch Wireless Power Model, April 2014 Matlab Code and documentation (Doc1, Doc2 and FAQ) at “http://members.greentouch.org/apps/org/workgroup/mobile/ \\
  documents.php?folder\_id=390”

\bibitem{ASM}      
F. E. Salem et al., ``Advanced Sleep Modes and Their Impact on Flow-Level Performance of 5G Networks," 2017 IEEE 86th Vehicular Technology Conference (VTC-Fall), Toronto, ON, 2017, pp. 1-7.

\bibitem{ref14}    
	www.freescale.com

\bibitem{ref15}    
	Phan-Huy, D.-T. et al, ``Dumb-to-perfect receiver throughput ratio maps of a time reversal wireless indoor system," in Proc. 20th International Conference on Telecommunications (ICT) 2013, Casablanca, 6-8 May 2013, pp.1,5
	
\bibitem{ref16}    	
	Phan-Huy, D.-T. et al, ``Adaptive large MISO downlink with Predictor Antenna array for very fast moving vehicles," 2013 International Conference on Connected Vehicles and Expo (ICCVE), pp.331,336, 2-6 Dec. 2013.
	
\bibitem{ref17}    
  Phan-Huy, D.-T. et al, ``Making 5G adaptive antennas work for very fast moving vehicles", in IEEE Intelligent Transportation Systems Magazine, vol. 7, no. 2, pp. 71-84, Summer 2015.

\bibitem{ref18}    
  Phan-Huy, D.-T. et al, ``Frequency Division Duplex Time Reversal," in Proc. IEEE Global Telecommunications Conference 2011 (GLOBECOM) 2011, Houston, 5-9 Dec. 2011, pp.1, 5

\bibitem{refA}    
Liu, V et al. ``Ambient Backscatter: Wireless Communication out of Thin Air", in Proc. SIGCOMM, 2013.

\bibitem{refB}    
Van Huynh, N. et al, ``Ambient Backscatter Communications: A Contemporary Survey," in IEEE Communications Surveys \& Tutorials, May 2018.

\bibitem{refC}    
Finkenzeller, K. ``RFID Handbook, Fundamentals and Applications in Contactless Smart Cards, Radio Frequency Identification and Near-Field Communication", Third Edition, Wiley, 2010.

\bibitem{refD}    
Lavric, A. et al, ``Internet of Things and LoRa™ Low-Power Wide-Area Networks: A survey," 2017 International Symposium on Signals, Circuits and Systems (ISSCS), Iasi, 2017, pp. 1-5. 

\bibitem{refE}    
3GPP TR 45.820 V13.1.0 (2015-11) Technical Report 3rd Generation Partnership Project; Technical Specification Group GSM/EDGE Radio Access Network; Cellular system support for ultra-low complexity and low throughput Internet of Things (CIoT) (Release 13).


\bibitem{ref19}    
  Ilab-Cairo study in 2011.

\bibitem{ref20}      
  UMF Specifications - release 3”, UNIVERSELF project Deliverable 2.4, 2013

\bibitem{ref21}    
  Autonomic network engineering for the self-managing Future Internet (AFI): GANA Architectural Reference Model for Autonomic Networking, Cognitive Networking and Self-Management, ETSI GS AFI 002. (2011).

\bibitem{ref22}      
Most of these DC are mixed telecom / IT sites, notably in smaller countries.

\bibitem{ref23}      
  R. Modrzejewski et. All ``Energy Efficient Content Distribution in an ISP Network" in proceedings of the 2013 IEEE Global Communications Conference, GLOBECOM 2013, Atlanta, GA, USA, December 9-13, 2013.

\bibitem{ref24}      
Y. Denis et al, Research paper 2013 - The Next Generation Central Offices

\bibitem{ref25}      
Denis Y. ``The Next Generation Central Offices a breakthrough evolution scenario for the optical access infrastructure", Orange research paper, April 2013.

\bibitem{ref26}      
GreenTouch Technical Solutions for Energy  Efficient Mobile Networks White  Paper, 2015.

\bibitem{ref27}      
J. Sekhar, G. Jeba ``Energy Efficient VM Live Migration in Cloud Data Centers" IJCSN International Journal of Computer Science and Network, Vol 2, Issue 2, April 2013. 

\bibitem{ref28}      
S. Le Masson, et al., ``Simplified Air-conditioning for telecommunication switches", Telecommunications Energy Conference, 2006, INTELEC '06, 28th Annual International, Oct. 2006.

\bibitem{ref29}      
  S. Le Masson, et al., ``New thermal architecture for future green data centres", Telecommunications Energy Conference (INTELEC), 32nd International; Jan. 2010.
\bibitem{ref30}      
  D. Nörtershäuser et al., ``A Step towards green datacenters: Enlarging climatic ranges – Studying the effects of the building", INTELEC San Diego 2008.
  
\bibitem{ref31}      
  S. Le Masson et al., ``Complex wall for indirect freecooling in datacenters", Telecommunications Energy Conference (INTELEC), 2012 IEEE 34th International, 01/2012
  
\bibitem{ref32}      
  S. Le Masson , ``Procede d'evacuation de chaleur degage a L’interieur d’un local par Ventilation a Debit Variable", French  patent 2 842 588, issued July 2002.
  
\bibitem{ref33}      
  D. Nörtershäuser et al., ``Exothermal building", International patent PCT/FR2010/052042, issued April 2011.
  
\bibitem{ref34}      
https://en.wikipedia.org/wiki/Fresnel\_lens.

\bibitem{RL}      
R. S. Sutton and A. G. Barto, ``Reinforcement learning: An introduction" MIT press Cambridge, 1998, vol. 1, no. 1.

\bibitem{qlearning}      
F. E. Salem et al., ``Reinforcement learning approach for Advanced Sleep Modes management in 5G networks," 2018 IEEE 87th Vehicular Technology Conference (VTC-Fall), Chicago, 2018, pp. 1-5.


\end{thebibliography}
\end{document}